\documentclass[11pt, a4paper]{article}
\usepackage[utf8]{inputenc}
\usepackage[T1]{fontenc}
\usepackage[margin=1in]{geometry}
\usepackage{amsmath,amssymb,amsfonts,mathtools}
\usepackage{graphicx}
\usepackage{booktabs}
\usepackage{array}
\usepackage{xcolor}
\usepackage[numbers,sort&compress]{natbib}
\usepackage{subcaption}
\usepackage{hyperref}
\usepackage{authblk}
\usepackage{lmodern}

\newcommand{\Hquad}{\hspace{1.8em}} 

\newcommand{\Bvec}{\vec{B}}
\newcommand{\xvec}{\vec{x}}
\newcommand{\grad}{\nabla}
\newcommand{\curl}{\nabla \times}
\newcommand{\divg}{\nabla \cdot}

\title{\textbf{Analytic non-axisymmetric 3D MHD equilibria without stellarator symmetry}}
\author[1,2]{Opal Issan}
\author[1]{Jonathan Citrin}
\author[3]{Matt Landreman}
\author[1]{Brendan Tracey}

\affil[1]{Google DeepMind, London, UK}
\affil[2]{Department of Mechanical and Aerospace Engineering, University of California San Diego, La Jolla, CA, USA}
\affil[3]{University of Maryland, College Park, MD, USA}

\date{\today}

\begin{document}
\maketitle

\begin{abstract}
Analytical solutions to the three-dimensional ideal magnetohydrodynamic equilibrium equations $(\curl\Bvec)\times\Bvec = \grad p$ and $\divg\Bvec = 0$ provide benchmarks for stellarator equilibrium codes.
Two families of non-axisymmetric solutions were recently discovered by~\citet{Landreman2026} as counterexamples to Grad's conjecture, but both satisfy stellarator symmetry.
Here, we generalize both families by introducing a symmetry-breaking parameter into each of the solutions, resulting in new non-axisymmetric solutions that break stellarator symmetry and reduce to the stellarator-symmetric solutions when the newly introduced parameter is set to zero.
\end{abstract}

\section{Introduction} \label{sec:intro}
In static ideal magnetohydrodynamics (MHD), an isotropic plasma pressure $p(\xvec)$ and magnetic field $\Bvec(\xvec) = (B_{x}, B_{y}, B_{z})$  satisfy
\begin{equation} \label{eq:mhd_system}
  (\curl\Bvec) \times \Bvec = \grad p, \qquad \divg\Bvec = 0, \qquad \Bvec \cdot \grad\psi = 0,
\end{equation}
where $p(\psi)$ is constant on smooth nested toroidal flux surfaces labeled by $\psi(\xvec) \in [0, \delta]$ with boundary flux-function value $\delta > 0$.
We denote Cartesian and cylindrical spatial coordinates by $\xvec \coloneq (x,y,z) \coloneq (R\cos\phi, R\sin\phi, z)$.
It is interesting to note that Eq.~\eqref{eq:mhd_system} is isomorphic to the steady incompressible Euler equations.
Using the vector identity $(\curl\Bvec)\times\Bvec = (\Bvec\cdot\grad)\Bvec - \frac{1}{2}\grad|\Bvec|^2$, when the magnetic tension is the gradient of a scalar potential $Q(\xvec)$, the force balance equation in~\eqref{eq:mhd_system} can be written as
\begin{equation} \label{eq:conservative_tension}
  (\Bvec \cdot \grad)\Bvec = -\grad Q, \qquad \grad p = -\grad\left(Q + \frac{|\Bvec|^2}{2}\right).
\end{equation} 

In axisymmetric geometry ($\partial_\phi = 0$), Eq.~\eqref{eq:mhd_system} reduces to the two-dimensional Grad--Shafranov equation, which has well-known analytical solutions, most notably the Solov'ev family and its extensions~\cite{Shafranov1966, Solovev1968, Zheng1996, CerfonFreidberg2010}.
Grad conjectured that smooth toroidal equilibria with non-constant pressure do not exist apart from specific configurations, such as axisymmetry or mirror-reflection symmetry with zero rotational transform ($\iota = 0$)~\cite{Grad1967, lortz1970existenz}.
Counterexamples to Grad's conjecture were found only very recently. 
In particular, \citet{GomezSerrano2026} proved the existence of non-axisymmetric equilibria with infinite rotational transform ($\iota = \infty$). 
Additionally, \citet{Landreman2026} constructed the first explicit closed-form equilibria with finite rotational transform, consisting of (1)~a constant-rotational-transform ($\iota = 2$) family obtained by stretching an axisymmetric Solov'ev equilibrium~\cite{Solovev1968} along the horizontal axes, and (2)~a sheared-rotational-transform ($\iota'(\psi) \neq 0$) family constructed in elliptic coordinates.

Both solutions in~\citet{Landreman2026} are \emph{stellarator symmetric}~\cite{DewarHudson1998}.
In Cartesian coordinates, stellarator symmetry is an invariance under a $\pi$-rotation about the $x$-axis, i.e.,
\begin{equation} \label{eq:stellarator_symmetry}
  \Bvec(\mathcal{S}\xvec) = -\mathcal{S} \Bvec(\xvec), \qquad \psi(\mathcal{S}\xvec) = \psi(\xvec), \qquad \mathcal{S} \coloneq \operatorname{diag}(1, -1, -1).
\end{equation}
Although the equilibria of~\citet{GomezSerrano2026} can break stellarator symmetry, they do not have an explicit closed-form expression, which limits their use for benchmarking MHD codes.
As noted in Section~4 of~\citet{Landreman2026}, constructing explicit analytical solutions that break stellarator symmetry has therefore remained an open problem.
From a practical standpoint, in three-dimensional equilibrium solvers such as VMEC~\cite{HirshmanWhitson1983}, SPEC~\cite{Hudson2012} or DESC~\cite{dudt2020desc}, stellarator symmetry halves the number of independent Fourier coefficients in the solution representation, so testing solver performance in non-stellarator-symmetric regimes requires benchmarks that break this symmetry.

In this paper, we construct two multi-parameter families of three-dimensional MHD equilibria that break stellarator symmetry and generalize the two solutions of~\citet{Landreman2026}.
Section~\ref{sec:family1} presents a constant-rotational-transform ($\iota = 2$) family obtained by adding an odd-in-$z$ term to the axisymmetric Solov'ev equilibrium~\cite{Solovev1968} before stretching it along the horizontal axes.
Section~\ref{sec:family2} presents a sheared-rotational-transform ($\iota'(\psi) \neq 0$) family in elliptic coordinates obtained by allowing the integration constant in the horizontal magnetic field to be complex, where the imaginary part acts as a symmetry-breaking phase shift.
Lastly, Section~\ref{sec:summary} discusses conclusions and future directions.

\section{Stretched asymmetric Solov'ev equilibrium with \texorpdfstring{$\iota = 2$}{iota=2}} \label{sec:family1}
We start by choosing two parameters: a horizontal elongation parameter $\epsilon \in (0, 1)$ and a vertical asymmetry parameter $\tau_{0} \in (-1, 1)$.
Let the coordinate stretch factors and auxiliary functions be
\begin{equation} \label{eq:f1_s_F}
  a \coloneq \sqrt{1 + \epsilon}, \Hquad b \coloneq \sqrt{1 - \epsilon}, \Hquad s \coloneq \frac{x^2}{a^2} + \frac{y^2}{b^2}, \Hquad F \coloneq \sqrt{1 - (1 - s)^2 - 4z^2 - 4\tau_{0} s z } > 0.
\end{equation}
The magnetic field is
\begin{equation} \label{eq:f1_B_cartesian}
  B_x = \frac{(2z + \tau_{0}  s)x - aF y/b}{s}, \qquad B_y = \frac{(2z +\tau_{0}  s)y + bFx/a}{s}, \qquad B_z = 1 - s - 2 \tau_{0}  z.
\end{equation}
With $\Omega \coloneq \sqrt{1 - \tau_{0}^2}$, the scalar function $Q$ in Eq.~\eqref{eq:conservative_tension} is
\begin{equation} \label{eq:f1_Q}
  Q = \frac{\Omega^2}{2}\,(x^2 + y^2 + 4z^2) + 2\tau_{0} z,
\end{equation}
and the flux function $\psi$ and plasma pressure $p$ satisfying Eq.~\eqref{eq:mhd_system} are
\begin{equation} \label{eq:f1_pressure_psi_cartesian}
  \psi = \frac{\Omega^2}{2} \left(Q + \frac{1}{2}|\Bvec|^2 - 1 + \frac{\tau_{0}^2}{2\Omega^2} + \frac{\Omega^2 \epsilon^2}{2} \right), \qquad p = p_a - \frac{2\psi}{\Omega^2},
\end{equation}
where $p_a$ is the pressure on the magnetic axis (i.e., $\psi = 0$).

Setting $\tau_{0}= 0$ in Eqs.~\eqref{eq:f1_s_F}--\eqref{eq:f1_pressure_psi_cartesian} recovers the stellarator-symmetric equilibrium of~\citet{Landreman2026}, whereas $\tau_{0} \neq 0$ breaks stellarator symmetry. 
\citet{Landreman2026} constructed the equilibrium by horizontally stretching an even-in-$z$ Solov'ev equilibrium~\cite{Solovev1968}, which preserves stellarator symmetry.
We break stellarator symmetry by adding an odd-in-$z$ term to the same Solov'ev equilibrium before applying the horizontal stretch in Eq.~\eqref{eq:f1_s_F}.

\subsection{Field-line coordinates} 
Let $\zeta \in \mathbb{R}$ parameterize a magnetic field line $\xvec(\zeta) = (x(\zeta), y(\zeta), z(\zeta))$ via
\begin{equation} \label{eq:f1_fieldline_ode}
  \Omega \frac{\mathrm{d}\xvec}{\mathrm{d}\zeta} = \Bvec\big(\xvec(\zeta)\big).
\end{equation}
Differentiating Eq.~\eqref{eq:f1_fieldline_ode} with respect to $\zeta$ gives $\Omega^2\,\mathrm{d}^2\xvec/\mathrm{d}\zeta^2 = (\Bvec\cdot\grad)\Bvec = -\grad Q$. Because $-\grad Q$ from Eq.~\eqref{eq:f1_Q} is linear in $\xvec$, every field line of Eq.~\eqref{eq:f1_B_cartesian} is a harmonic oscillator trajectory:
\begin{equation} \label{eq:f1_oscillator_odes}
  \frac{\mathrm{d}^2 x}{\mathrm{d}\zeta^2} = -x, \qquad \frac{\mathrm{d}^2 y}{\mathrm{d}\zeta^2} = -y, \qquad \frac{\mathrm{d}^2 z}{\mathrm{d}\zeta^2} = -4(z - z_c), \qquad z_c \coloneq -\frac{\tau_{0}}{2\Omega^2}.
\end{equation}
Thus $x(\zeta)$ and $y(\zeta)$ oscillate with frequency $1$, while $z(\zeta)$ oscillates about the vertical offset $z_c$ with frequency $2$.
Solving Eqs.~\eqref{eq:f1_fieldline_ode}--\eqref{eq:f1_oscillator_odes} introduces two field-line label coordinates $(u, v)$.
The field-line to Cartesian coordinate map $(u,v,\zeta) \mapsto (x,y,z)$ is
\begin{equation} \label{eq:f1_xyz_map}
\begin{aligned}
  x &= \frac{a}{\Omega}\left(L\cos(\zeta) + \frac{u\cos(\zeta) + v\sin(\zeta)}{L}\right), \\
  y &= \frac{b}{\Omega}\left(L\sin(\zeta) + \frac{v\cos(\zeta) - u\sin(\zeta)}{L}\right), \\
  z &= z_c + \frac{v\cos(2\zeta+\gamma) - u\sin(2\zeta+\gamma)}{\Omega^2}, \\
  \gamma &\coloneq \arcsin(\tau_{0}), \qquad L \coloneq \sqrt{\frac{1 + \sqrt{1 - 4(u^2 + v^2)}}{2}}.
\end{aligned}
\end{equation}
The only difference between Eq.~\eqref{eq:f1_xyz_map} and~\cite[Eq.~2.11]{Landreman2026} is that $(x, y)$ are scaled by $1/\Omega$, while $z$ is shifted by $z_{c}$, scaled by $1/\Omega^2$, and phase-shifted by $\gamma$.
Moreover, similar to~\citet{Landreman2026}, the coordinate Jacobian determinant is a global constant:
\begin{equation} \label{eq:f1_jacobian}
  \det\frac{\partial(x,y,z)}{\partial(u,v,\zeta)} = -\frac{ab}{\Omega^3} < 0.
\end{equation}

\subsection{Flux surfaces and magnetic axis} 
Using the field-line to Cartesian coordinate map in Eq.~\eqref{eq:f1_xyz_map}, we now express the flux function $\psi$ in field-line coordinates to identify the magnetic axis geometry.
In field-line coordinates, the flux function~\eqref{eq:f1_pressure_psi_cartesian} simplifies to
\begin{equation*} 
  \psi = (u - u_a)^2 + v^2, \qquad u_a \coloneq -\frac{\Omega^2\epsilon}{2}.
\end{equation*}
Surfaces of constant $\psi$ are circular cylinders of radius $\sqrt{\psi}$ centered on the magnetic axis $(u_a, 0, \zeta)$, which map to smooth nested tori in Cartesian space for any boundary flux $\delta$ satisfying $0 < \delta < (1 - \Omega^2\epsilon)^2/4$.
Increasing $|\tau_{0}|$ reduces $\Omega^2$, which shifts $u_a$ toward the origin and increases the upper bound of $\delta$ from $(1 - \epsilon)^2/4$ at $\tau_{0} = 0$ toward $1/4$ as $|\tau_{0}| \to 1$.

\begin{figure}
  \centering
  \begin{subfigure}[t]{0.32\textwidth}
    \centering
    \caption{3D magnetic axis}
    \includegraphics[width=\textwidth]{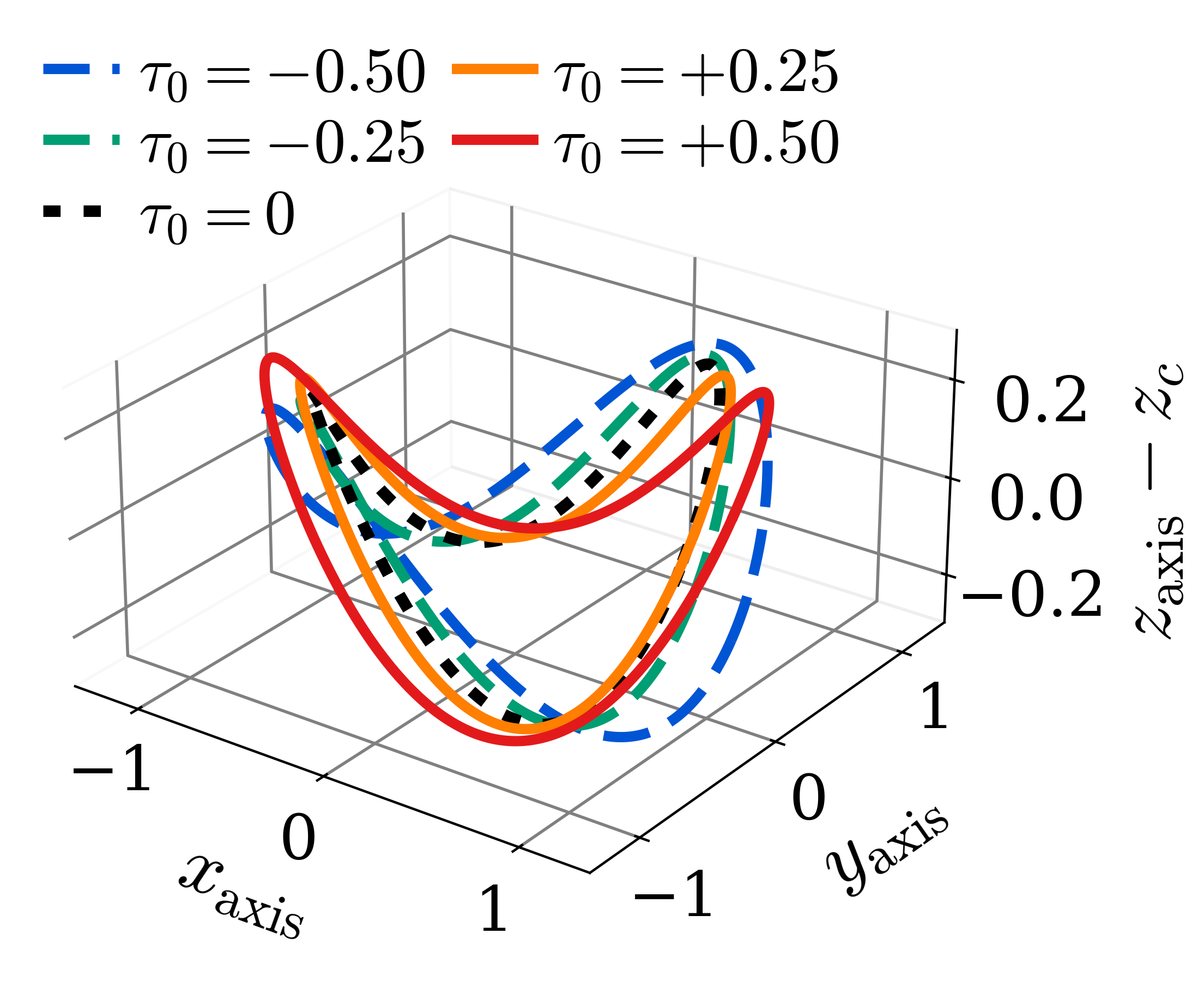}
    \label{fig:f1_3d_axes_all}
  \end{subfigure}
  \hfill
  \begin{subfigure}[t]{0.32\textwidth}
    \centering
    \caption{Horizontal projection}
    \includegraphics[width=\textwidth]{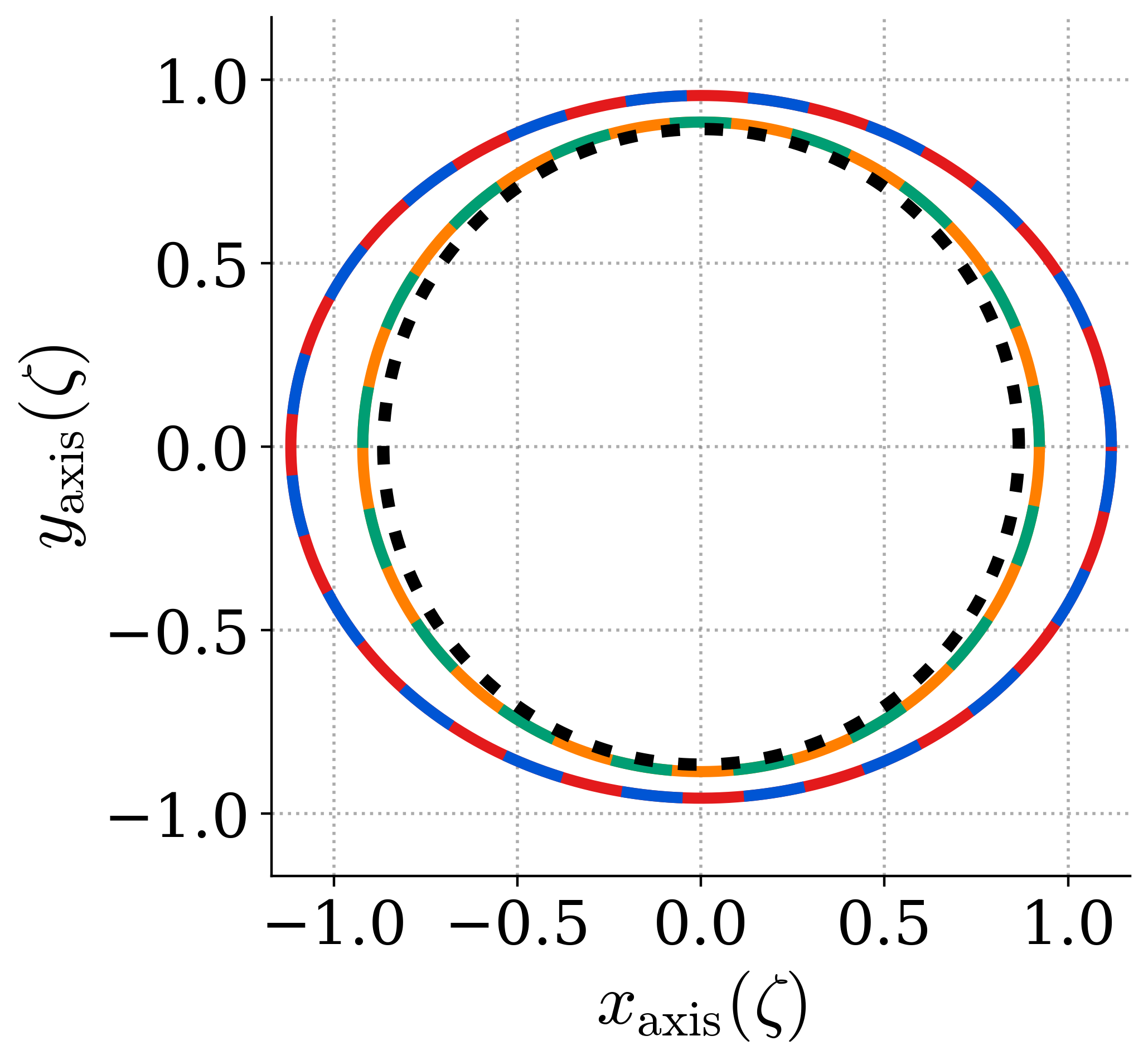}
    \label{fig:f1_axis_xy}
  \end{subfigure}
  \hfill
  \begin{subfigure}[t]{0.32\textwidth}
    \centering
    \caption{Vertical projection}
    \includegraphics[width=\textwidth]{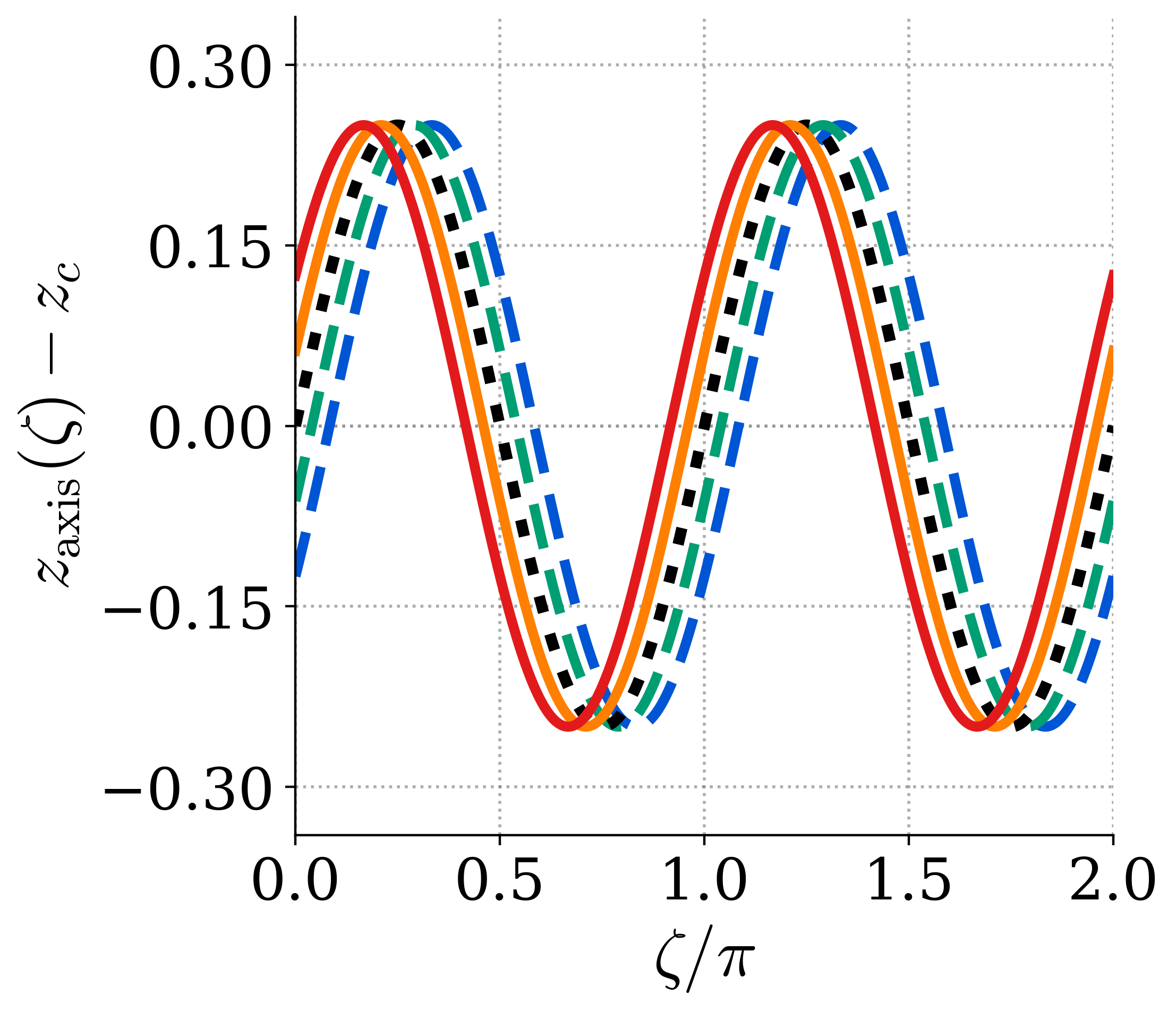}
    \label{fig:f1_axis_z}
  \end{subfigure}
  \caption{(a)~Three-dimensional magnetic axis curves for $\epsilon = 0.5$ and corresponding (b)~horizontal and (c)~vertical projections. Setting a nonzero $\tau_{0}$ results in a magnetic axis with an elliptic horizontal projection and a vertical phase shift as a function of the field-line coordinate $\zeta$.}
  \label{fig:family1_axis}
\end{figure}

Evaluating Eq.~\eqref{eq:f1_xyz_map} at $(u_a, 0, \zeta)$ gives the following magnetic axis curve
\begin{equation} \label{eq:f1_axis_curve}
  \xvec_{\mathrm{axis}}(\zeta) = \left(A_x \cos(\zeta), A_y \sin(\zeta), z_c + \frac{\epsilon}{2}\sin(2\zeta + \gamma)\right),
\end{equation}
with horizontal semi-axes
\begin{equation*}
  A_x \coloneq \frac{\sqrt{(1+\epsilon)(1 - \Omega^2\epsilon)}}{\Omega}, \qquad A_y \coloneq \frac{\sqrt{(1-\epsilon)(1 + \Omega^2\epsilon)}}{\Omega}.
\end{equation*}
Figure~\ref{fig:family1_axis} shows the magnetic axis geometry for various values of $\tau_{0}$.
Figure~\ref{fig:f1_3d_axes_all} illustrates that the three-dimensional magnetic axis retains a saddle shape that is radially expanded and toroidally twisted by $\tau_{0}$.
The horizontal projection of the magnetic axis shown in Figure~\ref{fig:f1_axis_xy} is a circle of radius $\sqrt{1-\epsilon^2}$ when $\tau_{0} = 0$ and an ellipse with $A_x > A_y$ when $\tau_{0} \neq 0$.
The horizontal projection is invariant to the sign of $\tau_{0}$ because the horizontal components of Eq.~\eqref{eq:f1_axis_curve} only depend on $\tau_{0}^2$ via $\Omega$.
Furthermore, Figure~\ref{fig:f1_axis_z} shows that $\tau_{0}$ has a simple impact on the vertical profile $z_{\mathrm{axis}}(\zeta) - z_c = \frac{\epsilon}{2}\sin(2\zeta + \gamma)$, introducing a $\gamma$ phase shift.

\begin{figure}
  \centering
  \begin{subfigure}{0.32\textwidth}
    \centering
    \caption{$\tau_{0} = -0.5$}
    \includegraphics[width=\textwidth]{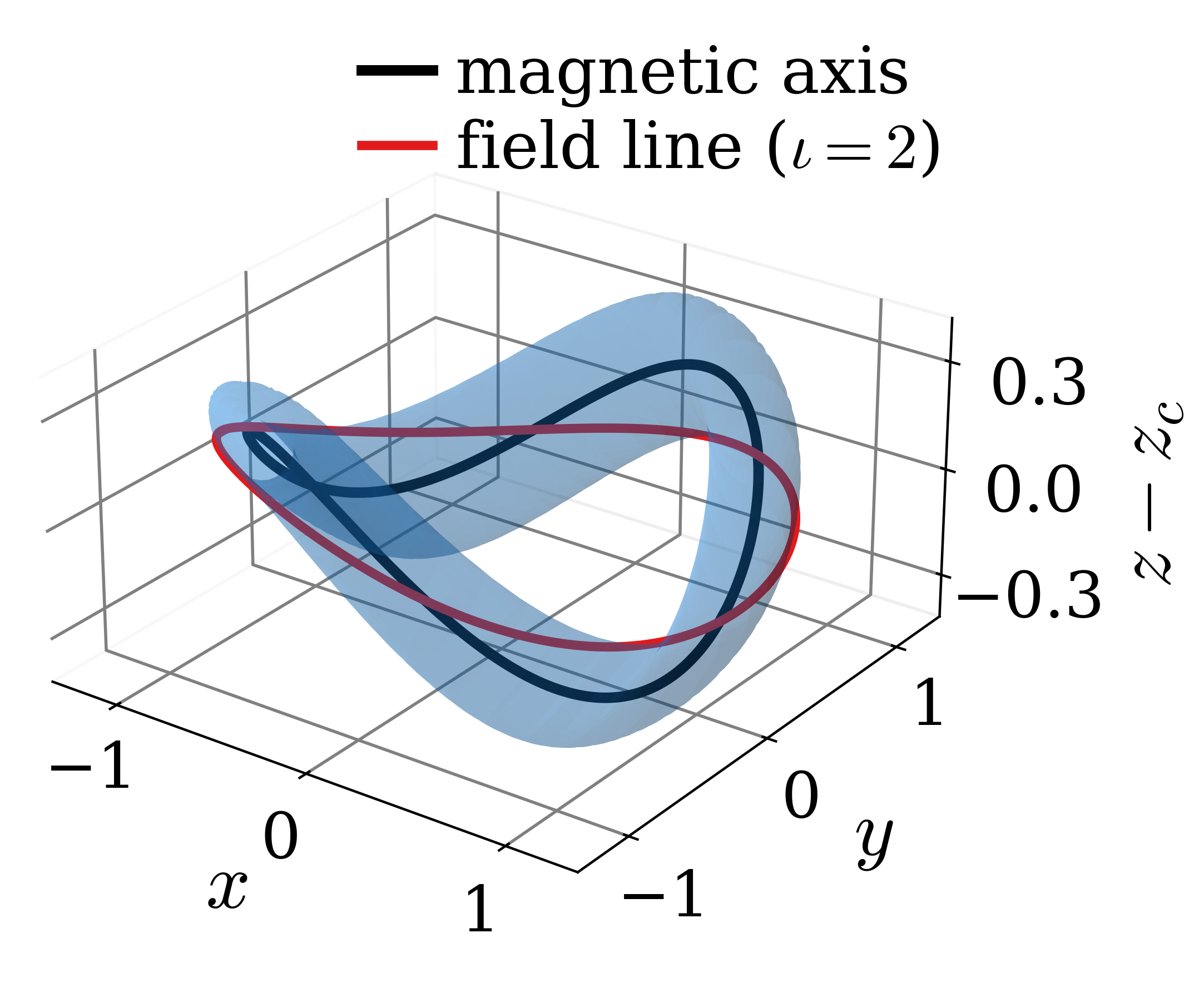}
  \end{subfigure}
  \hfill
  \begin{subfigure}{0.32\textwidth}
    \centering
    \caption{$\tau_{0} = 0$}
    \includegraphics[width=\textwidth]{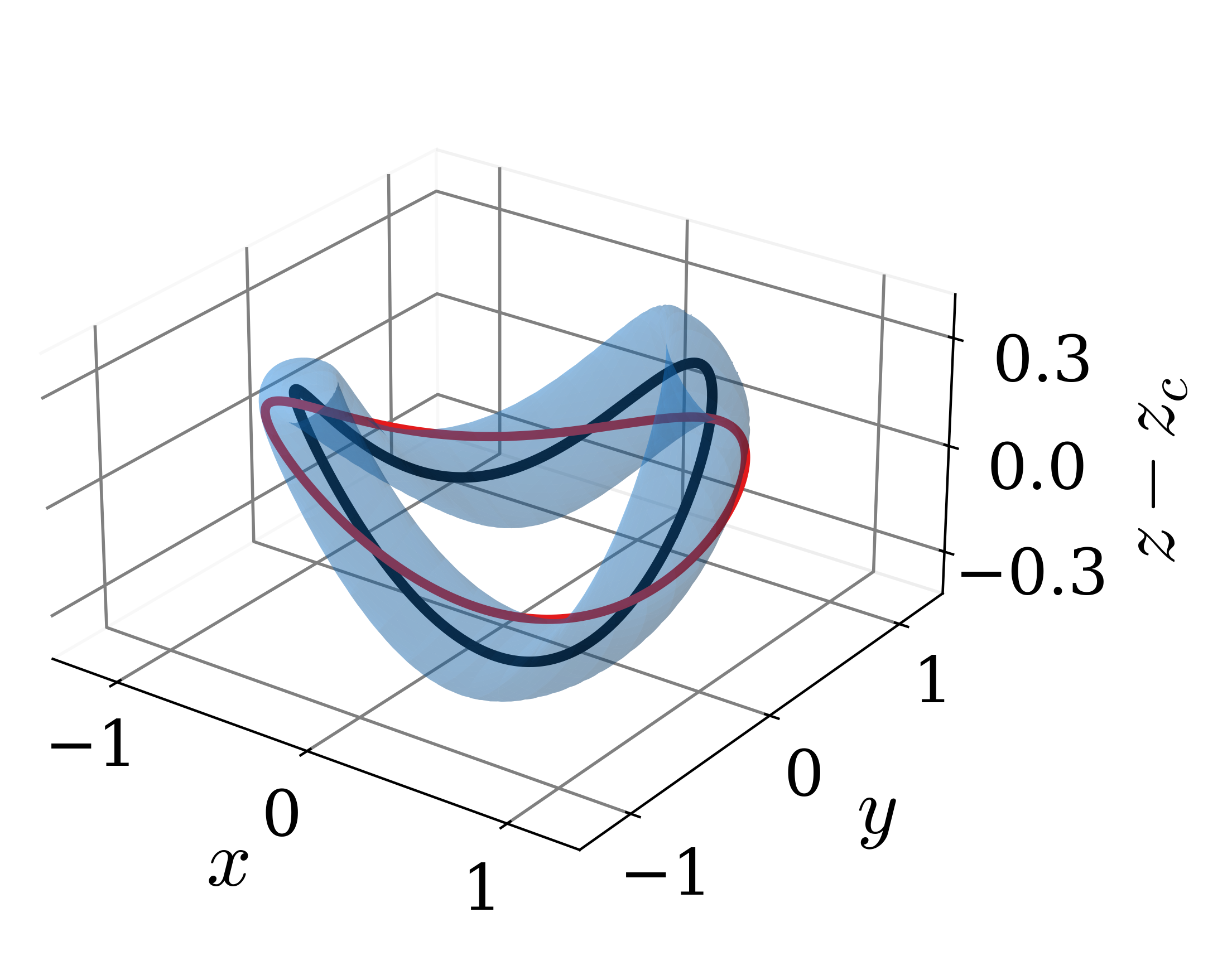}
  \end{subfigure}
  \hfill
  \begin{subfigure}{0.32\textwidth}
    \centering
    \caption{$\tau_{0} = 0.5$}
    \includegraphics[width=\textwidth]{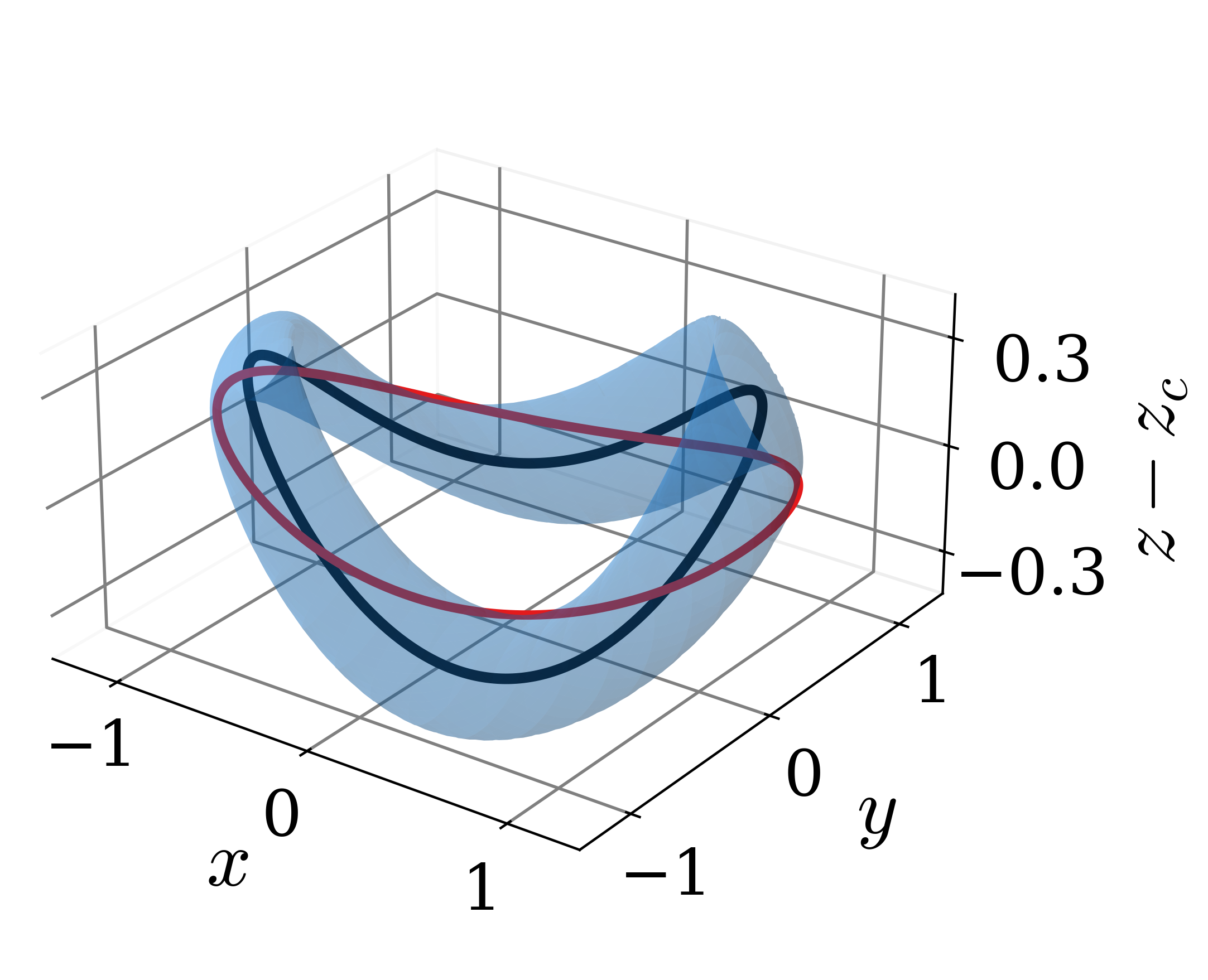}
  \end{subfigure}
  \caption{Three-dimensional boundary flux surface $\psi = \delta = 1/64$, magnetic axis, and a closed field line for $\epsilon = 0.5$ with (a)~$\tau_{0} = -0.5$, (b)~$\tau_{0} = 0$, and (c)~$\tau_{0} = 0.5$.}
  \label{fig:family1_3d}
\end{figure}

Surrounding the magnetic axis, the boundary flux surface $\psi = \delta = 1/64$ and a closed field line are shown in Figure~\ref{fig:family1_3d}. 
Following the analysis in~\citet{Landreman2026}, the $2\pi$-periodicity of $(x,y)$ and $2\zeta$ poloidal winding in Eq.~\eqref{eq:f1_xyz_map} yield a uniform rotational transform $\iota(\psi) = 2$.
Increasing $|\tau_{0}|$ reduces $\Omega$, which expands the torus radially by $1/\Omega$ and thickens its poloidal cross-section vertically by $1/\Omega^2$ (see Eq.~\eqref{eq:f1_xyz_map}). 
Additionally, the phase shift $\gamma$ twists the vertical crests and troughs in opposite toroidal directions for $\tau_{0} = 0.5$ and $\tau_{0} = -0.5$.

\subsection{Rotational transform, volume, and volume-averaged \texorpdfstring{$\beta$}{beta}}
To avoid singular parallel currents on a rational flux surface with non-zero pressure gradient, an equilibrium must satisfy the Hamada condition~\cite{Hamada1962}, which requires the line integral $\oint \mathrm{d}\ell/|\Bvec|$ (where $\ell$ is arc length along a field line) to be identical for all closed field lines on that surface. Here, because $\iota(\psi) = 2$ is rational, every field line closes after one toroidal transit $\Delta\zeta = 2\pi$ (and two poloidal transits). From Eq.~\eqref{eq:f1_fieldline_ode}, $\mathrm{d}\ell = |\mathrm{d}\xvec/\mathrm{d}\zeta|\,\mathrm{d}\zeta = (|\Bvec|/\Omega)\,\mathrm{d}\zeta$, so $\mathrm{d}\ell/|\Bvec| = \mathrm{d}\zeta/\Omega$ and the closed-loop integral $\oint \mathrm{d}\ell/|\Bvec| = \int_0^{2\pi} \mathrm{d}\zeta/\Omega = 2\pi/\Omega$ is identical on all closed field lines, satisfying the Hamada condition.

\begin{figure}
  \centering
  \begin{subfigure}{0.45\textwidth}
    \centering
    \caption{Enclosed volume $V(\delta)$}
    \includegraphics[width=\textwidth]{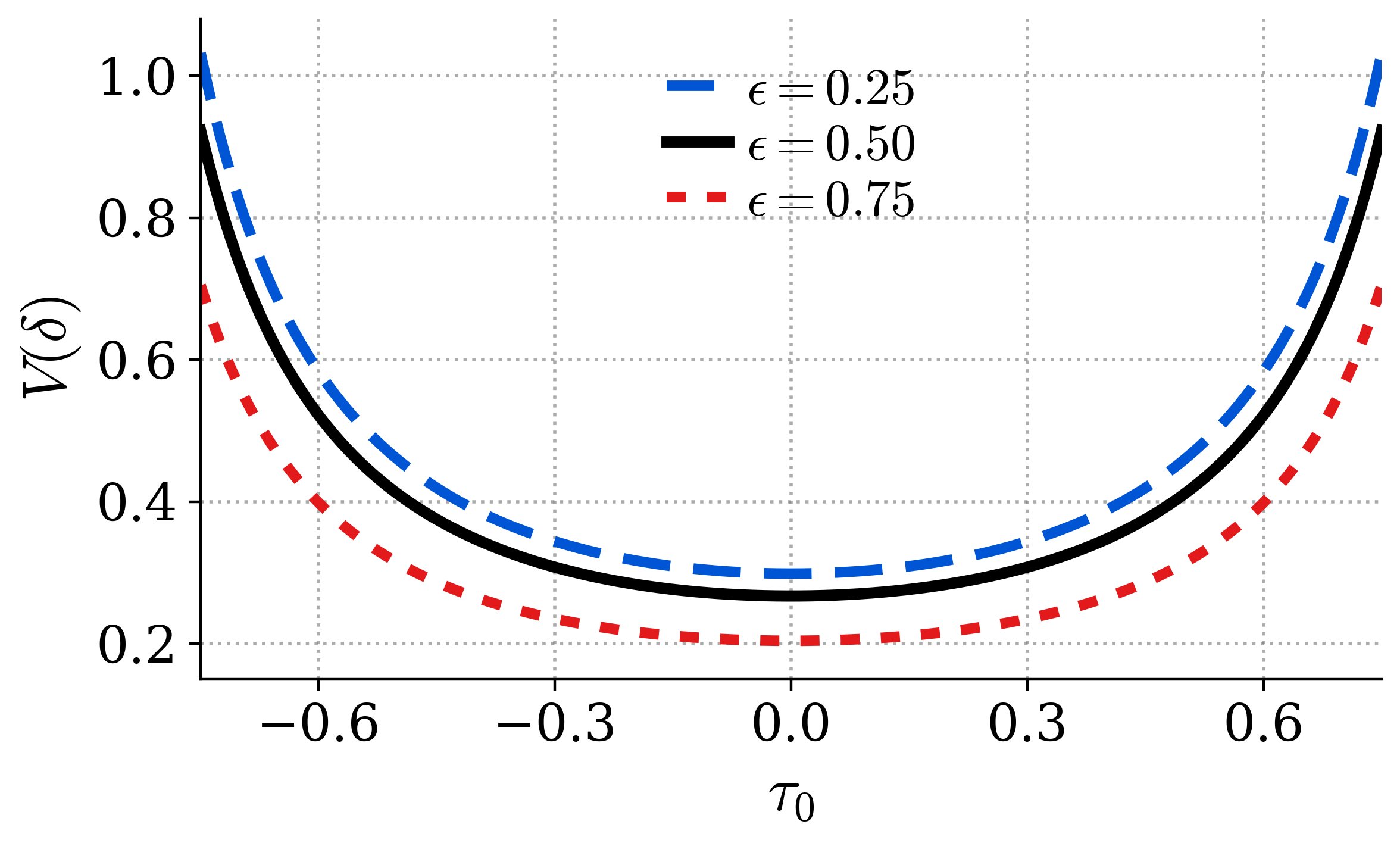}
    \label{fig:f1_volume_alpha}
  \end{subfigure}
  \hfill
  \begin{subfigure}{0.45\textwidth}
    \centering
    \caption{Volume-averaged $\beta_V$}
    \includegraphics[width=\textwidth]{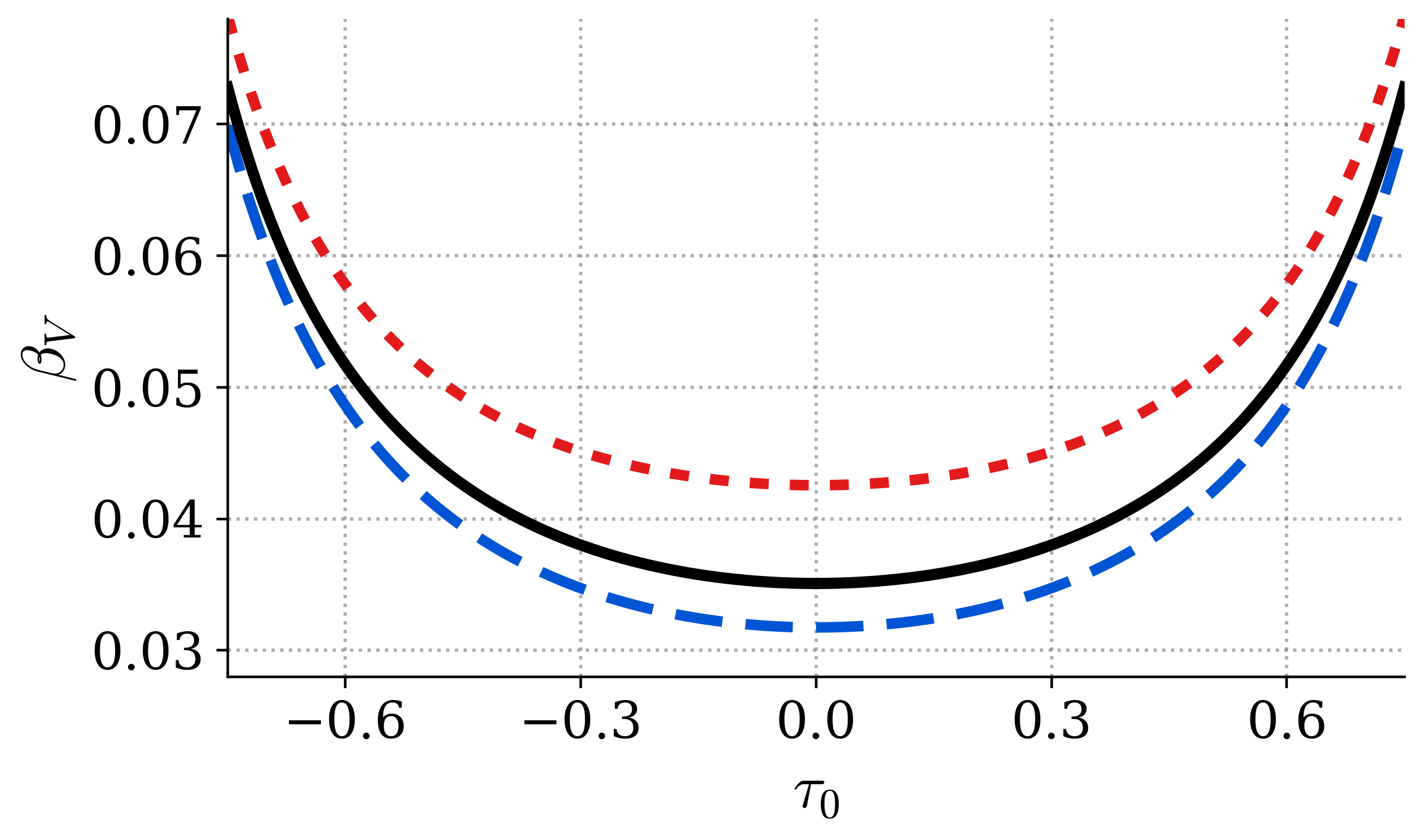}
    \label{fig:f1_beta_alpha}
  \end{subfigure}
  \caption{Enclosed volume $V(\delta)$ and volume-averaged plasma beta $\beta_V$ (with boundary pressure $p_b = 0$) vs. $\tau_{0}$ for the $\iota = 2$ equilibrium family with $\delta = 1/64$. Both quantities are even in $\tau_{0}$ and increase monotonically with $|\tau_{0}|$, whereas increasing $\epsilon$ reduces $V(\delta)$ and increases $\beta_V$.}
  \label{fig:family1_vol_beta}
\end{figure}

Integrating the constant Jacobian~\eqref{eq:f1_jacobian} gives the enclosed volume
\begin{equation} \label{eq:f1_iota_volume_flux}
    V(\psi) = \frac{2\pi^2 \sqrt{1-\epsilon^2} \psi}{\Omega^3}.
\end{equation}
With volume average $\langle f \rangle_V \coloneq \frac{1}{V(\delta)} \int_{0 \leq \psi \leq \delta} f \mathrm{d}V$ and boundary pressure $p_b \coloneq p(\delta) =  p_a - 2\delta/\Omega^2$, the volume-averaged plasma beta is
\begin{equation*} 
  \beta_V \coloneq \frac{2\langle p \rangle_V}{\langle |\Bvec|^2 \rangle_V} = \frac{2\big(p_b + \delta/\Omega^2\big)}{1 - \Omega^2\epsilon^2/2 + \delta/\Omega^2}.
\end{equation*}
Because $V(\delta)$ and $\beta_V$ depend on $\tau_{0}$ only through $\Omega$, both are even in $\tau_{0}$, as illustrated in Figure~\ref{fig:family1_vol_beta}.
From Eq.~\eqref{eq:f1_iota_volume_flux}, the enclosed volume scales as $V(\delta) \propto \sqrt{1 - \epsilon^2}/\Omega^{3}$, so it decreases with $\epsilon$ but increases with $|\tau_{0}|$ (as $\Omega$ decreases) as shown in Figure~\ref{fig:f1_volume_alpha}.
In contrast, for $p_b = 0$, $\beta_V$ increases monotonically with both $\epsilon$ and $|\tau_{0}|$ as shown in Figure~\ref{fig:f1_beta_alpha}.

\subsection{Broken stellarator symmetry} 
For $\tau_{0} \in (-1,1)\setminus\{0\}$, the equilibrium in Eqs.~\eqref{eq:f1_s_F}--\eqref{eq:f1_pressure_psi_cartesian} breaks stellarator symmetry~\eqref{eq:stellarator_symmetry}, as the linear-in-$z$ terms reverse sign under $\mathcal{S}\xvec = (x, -y, -z)$:
\begin{equation*}
  B_z(\mathcal{S}\xvec) = 1 - s + 2\tau_{0} z \neq 1 - s - 2 \tau_{0}  z =  B_z(\xvec).
\end{equation*}
In cylindrical coordinates, stellarator symmetry maps $(R, \phi, z) \mapsto (R, -\phi, -z)$, while the two-field-period ($N_{\mathrm{fp}} = 2$) discrete rotational symmetry about the $z$-axis maps $(R, \phi, z) \mapsto (R, \phi + \pi, z)$, so a stellarator-symmetric flux function must obey
\begin{equation} \label{eq:f1_rz_symmetry} 
  \psi(R, \phi, z) = \psi(R, -\phi, -z) = \psi(R, \pi - \phi, -z).
\end{equation}
Setting $\phi = 0$ and $\phi = \pi/2$ yields $\psi(R, 0, -z) = \psi(R, 0, z)$ and $\psi(R, \pi/2, -z) = \psi(R, \pi/2, z)$, requiring both cross-sections to be symmetric across $z = 0$, whereas setting $\phi = \pi/4$ yields $\psi(R, 3\pi/4, -z) = \psi(R, \pi/4, z)$, requiring the $\phi = \pi/4$ and $\phi = 3\pi/4$ cross-sections to be vertical reflections of one another across $z = 0$.

\begin{figure}[htbp]
  \centering
    \begin{subfigure}[t]{0.32\textwidth}
    \centering
    \caption{$\tau_{0} = 0$}
    \includegraphics[width=\textwidth]{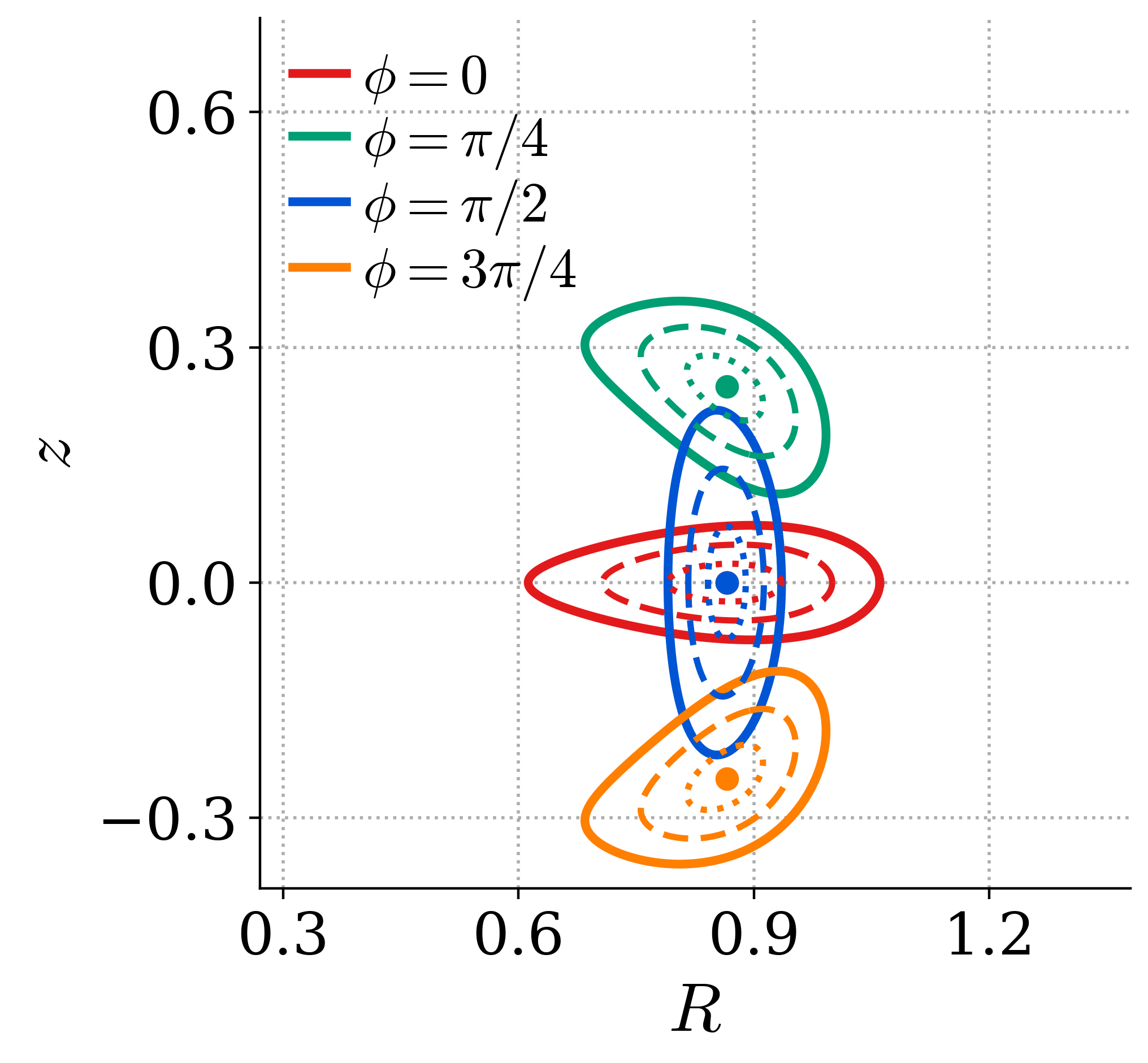}
    \label{fig:f1_rz_alpha_0}
  \end{subfigure}
  \begin{subfigure}[t]{0.32\textwidth}
    \centering
    \caption{$\tau_{0} = -0.5$}
    \includegraphics[width=\textwidth]{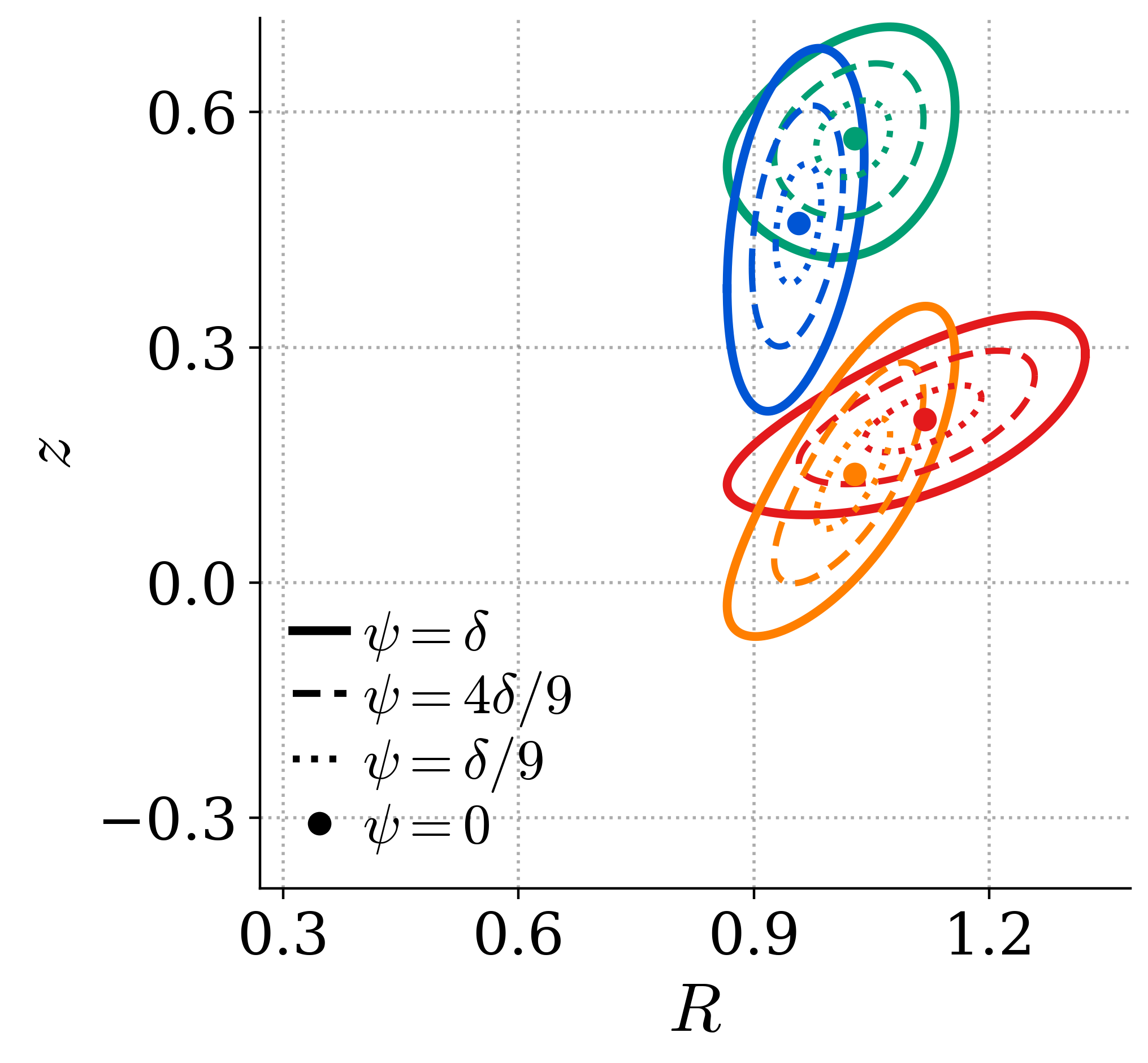}
    \label{fig:f1_rz_alpha_m050}
  \end{subfigure}
  \begin{subfigure}[t]{0.32\textwidth}
    \centering
    \caption{$\tau_{0} = 0.5$}
    \includegraphics[width=\textwidth]{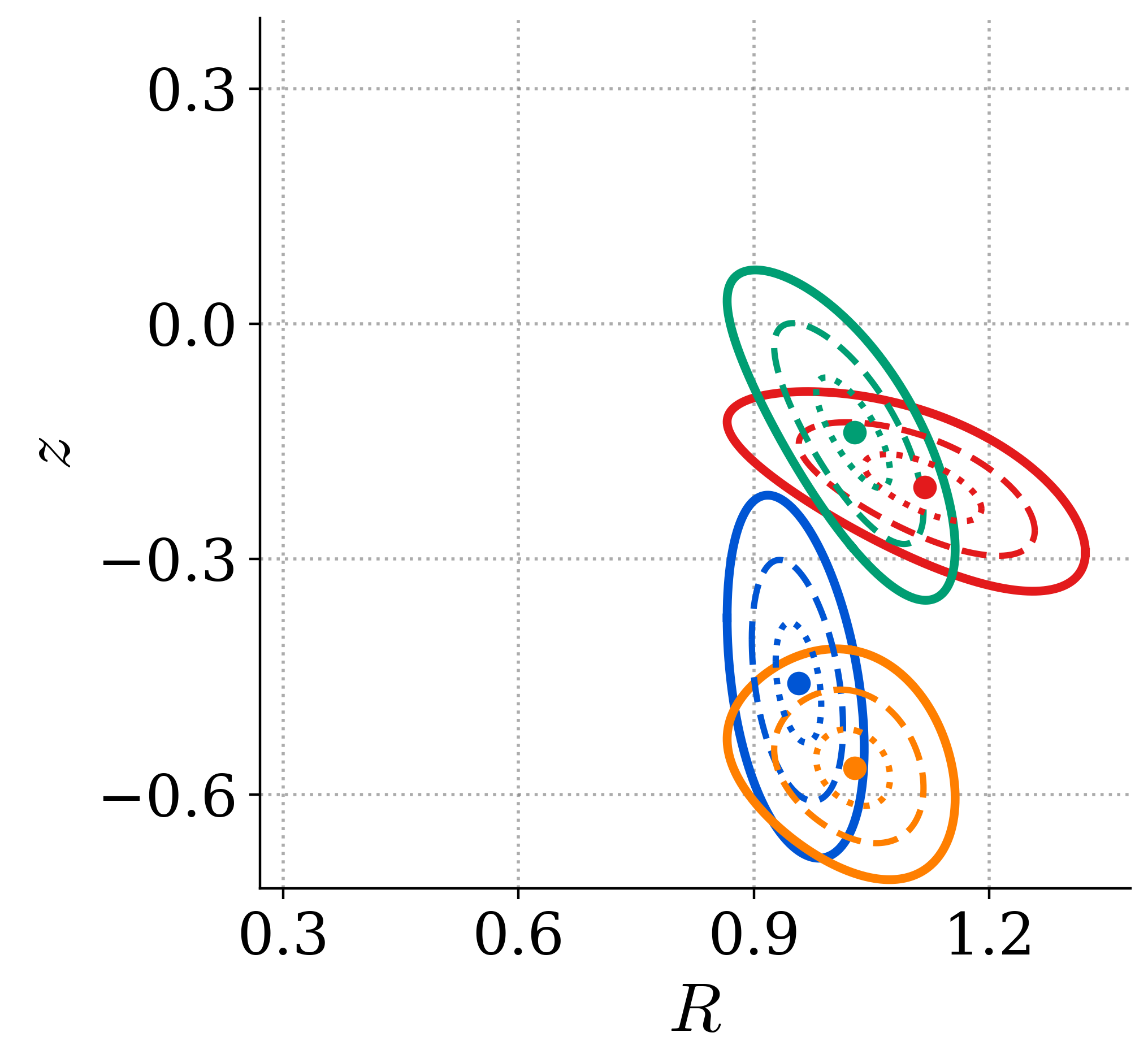}
    \label{fig:f1_rz_alpha_p050}
  \end{subfigure}
    \caption{Poloidal cross-sections for the $\iota = 2$ equilibrium family. Subfigure~(a) with $\tau_{0} = 0$ is stellarator symmetric, with  symmetric cross-sections at $\phi = 0, \pi/2$ and mirror-symmetric pairs at $\phi = \pi/4, 3\pi/4$, whereas nonzero $\tau_{0}$ breaks both symmetries.}
  \label{fig:family1_rz}
\end{figure}

Figure~\ref{fig:family1_rz} illustrates these poloidal cross-sections for $\epsilon = 0.5$ and $\delta = 1/64$.
Figure~\ref{fig:f1_rz_alpha_0} shows $\tau_{0}=0$ reproducing \cite[Fig.~1]{Landreman2026}, where the $\phi = 0$ and $\phi = \pi/2$ cross-sections are centered on and symmetric across $z = 0$, while the $\phi = \pi/4$ and $\phi = 3\pi/4$ slices are vertical reflections across $z = 0$.
In contrast, Figures~\ref{fig:f1_rz_alpha_m050}--\ref{fig:f1_rz_alpha_p050} show that when $\tau_{0} \neq 0$, the $\phi = 0$ and $\phi = \pi/2$ cross-sections separate radially and vertically while tilting diagonally, and the $\phi = \pi/4$ and $\phi = 3\pi/4$ slices lose their mirror symmetry.

\section{Complex-phase equilibrium with sheared \texorpdfstring{$\iota(\psi)$}{iota(psi)}} \label{sec:family2}
The second family of equilibria extends the solution by~\citet{Landreman2026} with sheared rotational transform profile ($\iota'(\psi) \neq 0$) and is parameterized by four scalars: the focal distance parameter $\epsilon > 0$, the radial axis parameter $S > 0$, the vertical scale parameter $\lambda > 0$, and the stellarator symmetry-breaking parameter $\tau_{1} \in \mathbb{R}$.
In Cartesian coordinates, we define the complex horizontal coordinates and auxiliary functions as
\begin{equation} \label{eq:f2_w_K_W_Xi_def}
  \omega \coloneq x + iy, \qquad K \coloneq \bar{\omega}\sqrt{1 + \frac{\epsilon}{\bar{\omega}^2}}, \qquad W \coloneq \frac{i\sin(\Xi)}{2K},  \qquad  \Xi \coloneq \omega K + \frac{\pi}{2} - S + i\tau_{1}, 
\end{equation}
where $\bar{\omega} \coloneq x - iy$ is the complex conjugate of $\omega$ and the square root in $K$ is chosen to have positive real part.
The magnetic field is
\begin{equation} \label{eq:f2_B_cartesian}
  B_x + iB_y = e^{-i\lambda z}W, \qquad
  B_z = \frac{1}{\lambda}\operatorname{Re} \left(e^{-i\lambda z}\cos(\Xi)\right).
\end{equation}
The scalar function $Q$, flux function $\psi$, and plasma pressure $p$ satisfying Eqs.~\eqref{eq:mhd_system}--\eqref{eq:conservative_tension} are
\begin{equation} \label{eq:f2_Q_pressure_psi}
  Q = -\frac{|W|^2}{2} + \frac{\sin^2(\lambda z)}{2\lambda^2}, \qquad   \psi = \frac{\sin^2(\lambda z) + \lambda^{2} B_z^{2}}{2}, \qquad p = p_a - \frac{\psi}{\lambda^2}. 
\end{equation}

Setting $\tau_{1} = 0$ in Eq.~\eqref{eq:f2_w_K_W_Xi_def} recovers the stellarator-symmetric equilibrium of~\citet{Landreman2026}, whereas $\tau_{1} \neq 0$ breaks stellarator symmetry.
\citet{Landreman2026} assumed a magnetic field that varies sinusoidally in the vertical direction, reducing the three-dimensional force balance in $(x, y, z)$ to an analytically solvable two-dimensional differential equation in $(x, y)$ for $W$ (defined in Eq.~\eqref{eq:f2_w_K_W_Xi_def}).
Solving this two-dimensional differential equation leaves a constant of integration.
Whereas~\citet{Landreman2026} took this constant to be real, we allow it to be complex, and its imaginary part breaks stellarator symmetry.

\subsection{Elliptic coordinates, flux surfaces, and magnetic axis} 
To obtain an explicit parameterization of the flux surfaces and magnetic axis, we follow~\citet{Landreman2026} and introduce elliptic coordinates $(\sigma, \zeta) \in (0, \infty) \times [0, 2\pi)$ in the horizontal plane:
\begin{equation} \label{eq:f2_xy}
  x = a_c \cos(\zeta), \Hquad y = b_c \sin(\zeta), \Hquad 
a_c \coloneq \sqrt{\frac{h - \epsilon}{2}},  \Hquad b_c \coloneq \sqrt{\frac{h + \epsilon}{2}}, \Hquad h \coloneq \sqrt{4\sigma^2 + \epsilon^2}.
\end{equation}
Because $\psi$ is a sum of two squares, see Eq.~\eqref{eq:f2_Q_pressure_psi}, each flux surface becomes a circle $\psi = (X^2 + Y^2)/2$ in 
\begin{equation} \label{eq:f2_X_def}
  X \coloneq \lambda B_z = -\sin(\sigma - S)\cosh(\tilde{\nu}) \cos(\lambda z) - \cos(\sigma - S)\sinh(\tilde{\nu}) \sin(\lambda z), \Hquad Y \coloneq -\sin(\lambda z),
\end{equation}
with $\tilde{\nu} \coloneq \epsilon\sin(2\zeta)/2 + \tau_{1}$. 
Together with Eq.~\eqref{eq:f2_xy}, this defines a smooth, invertible coordinate map $(X, Y, \zeta) \mapsto (x, y, z)$ with the following Jacobian determinant
\begin{equation*}
  \det\frac{\partial(x,y,z)}{\partial(X,Y,\zeta)} = -\frac{|K|^2}{\lambda h\sqrt{\big(\cosh^2(\tilde{\nu}) - 2\psi\big)(1 - Y^2)}} < 0.
\end{equation*}
The surfaces of constant $\psi$ map to smooth nested tori in Cartesian space for any boundary flux $\delta$ satisfying
\begin{equation*}
  0 < \delta < \frac{1}{2}\sin^2\big(\min(S, \pi/2)\big).
\end{equation*}

Setting $\psi = 0$ ($X = Y = 0$) in Eq.~\eqref{eq:f2_X_def} forces $\sigma = S$ and $z = 0$ for every $\tau_{1} \in \mathbb{R}$. Unlike Section~\ref{sec:family1}, where the newly introduced parameter $\tau_{0}$ deforms the magnetic axis, here the magnetic axis is independent of $\tau_{1}$, tracing the same planar ellipse as~\citet{Landreman2026}:
\begin{equation} \label{eq:f2_magnetic_axis}
  \xvec_{\mathrm{axis}}(\zeta) = \big(a_c \cos(\zeta), b_c \sin(\zeta), 0\big),\qquad a_{c} =  \sqrt{\frac{\sqrt{4S^2 + \epsilon^2} -\epsilon}{2}}, \qquad  b_{c} = \sqrt{\frac{\sqrt{4S^2 + \epsilon^2} + \epsilon}{2}}.
\end{equation}
As seen from Eq.~\eqref{eq:f2_magnetic_axis}, the magnetic axis is independent of both $\tau_{1}$ and $\lambda$, and is only controlled by $\epsilon$ and $S$.
Increasing $\epsilon$ deforms the magnetic axis from a circle ($\epsilon = 0$) into an elongated ellipse along the $y$-axis while preserving the enclosed area $\pi a_c b_c = \pi S$, whereas varying $S$ traces a nested family of ellipses sharing the foci $(0, \pm\sqrt{\epsilon})$.

\begin{figure}[htbp]
  \centering
  \begin{subfigure}{0.32\textwidth}
    \centering
    \caption{$\tau_{1} = -0.40$}
    \includegraphics[width=\textwidth]{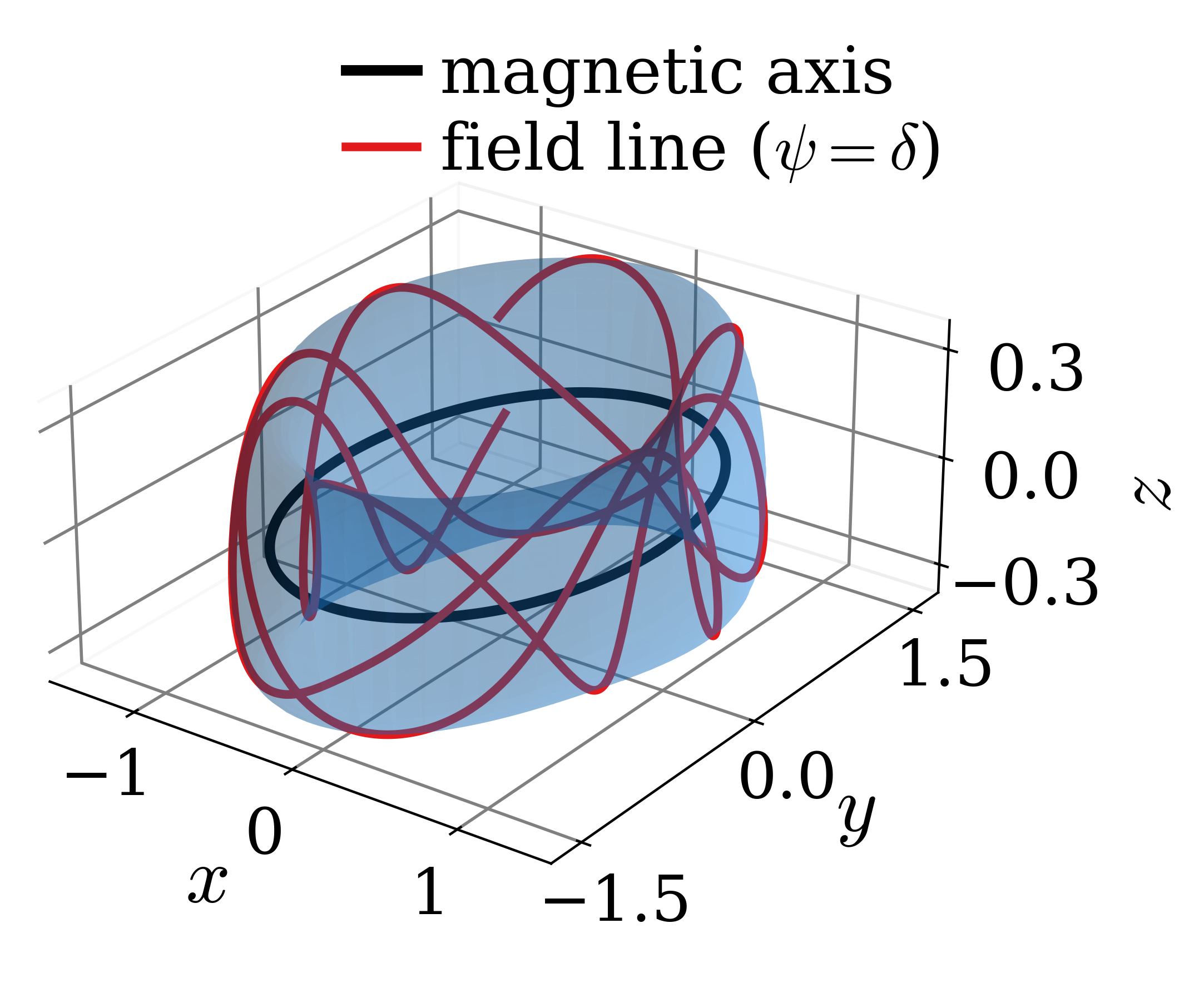}
    \label{fig:f2_3d_m040}
  \end{subfigure}
  \begin{subfigure}{0.32\textwidth}
    \centering
    \caption{$\tau_{1} = 0$}
    \includegraphics[width=\textwidth]{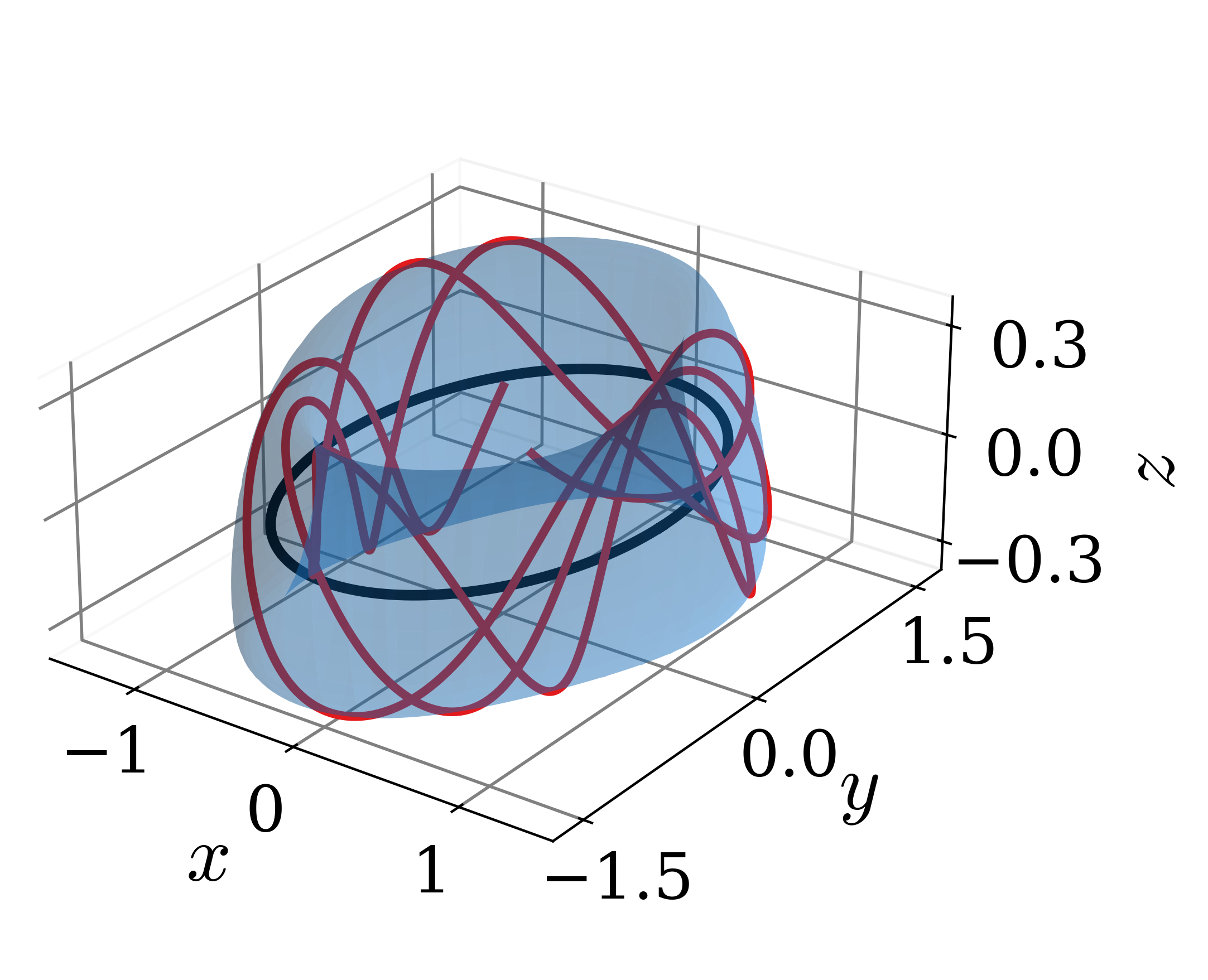}
    \label{fig:f2_3d_0}
  \end{subfigure}
  \begin{subfigure}{0.32\textwidth}
    \centering
    \caption{$\tau_{1} = 0.40$}
    \includegraphics[width=\textwidth]{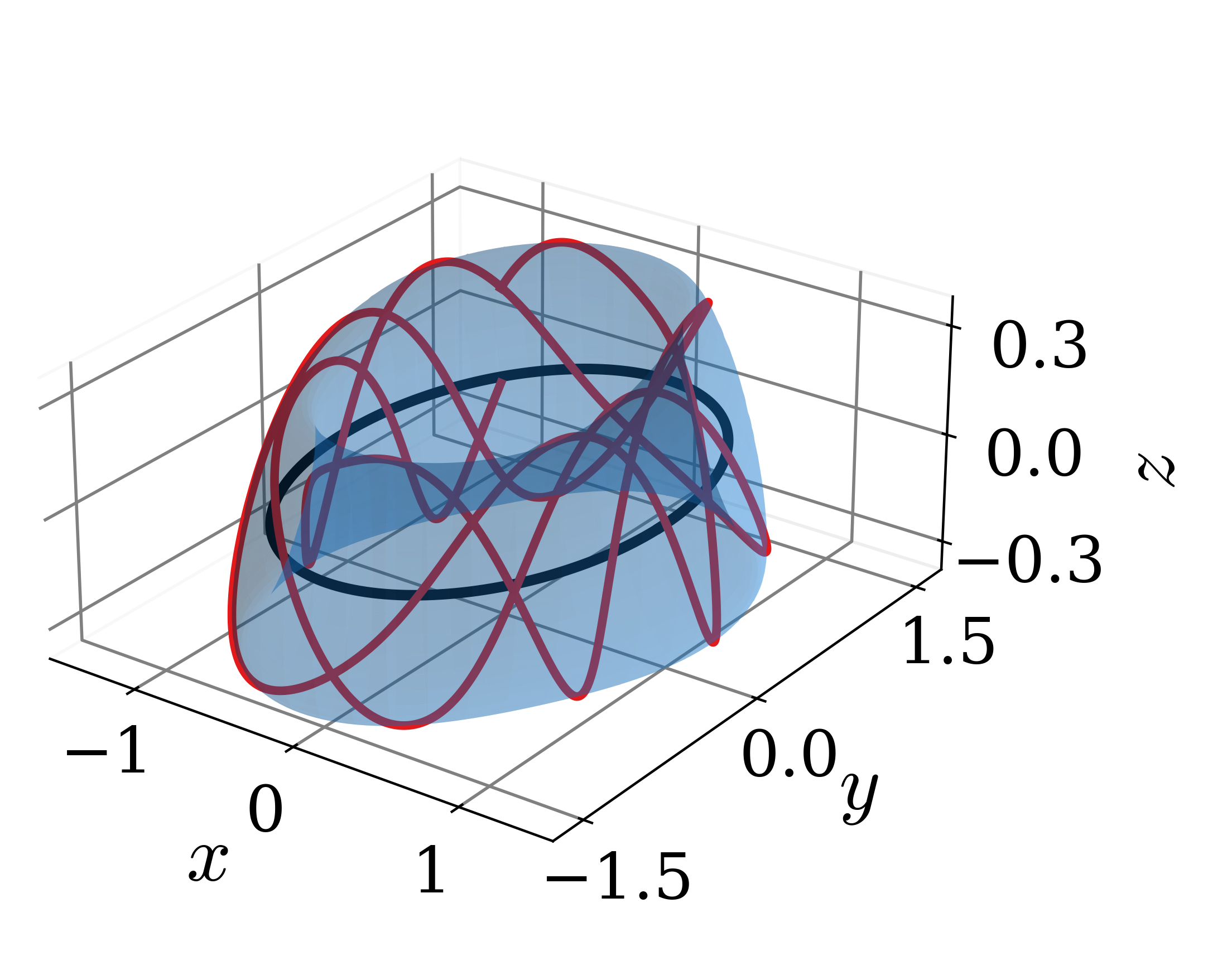}
    \label{fig:f2_3d_p040}
  \end{subfigure}
  \caption{Three-dimensional boundary flux surfaces $\psi = \delta = 0.14$, planar elliptical magnetic axis $\psi = 0$, and a field line on $\psi = \delta$ traced over three toroidal transits for the sheared-$\iota$ equilibrium family. }
  \label{fig:family2_3d}
\end{figure}

Figure~\ref{fig:family2_3d} shows the three-dimensional boundary flux surface $\psi = \delta = 0.14$ and a field line traced over three toroidal transits ($\Delta\zeta = 6\pi$) for $\epsilon = 1.2$, $S = 1.4$, $\lambda = 1.6$, for various values of $\tau_{1}$. 
Across all three panels, the elliptical magnetic axis remains fixed, as it is independent of $\tau_{1}$.
The difference between the traced field lines suggests that the rotational transform depends on $\tau_{1}$, which we derive in more detail in Section~\ref{subsec:f2_iota_beta}.

\subsection{Rotational transform} \label{subsec:f2_iota_beta}
\begin{figure}[htbp]
  \centering
  \begin{subfigure}[b]{0.49\textwidth}
    \centering
    \caption{$\tau_{1} \le 0$}
    \includegraphics[width=\textwidth]{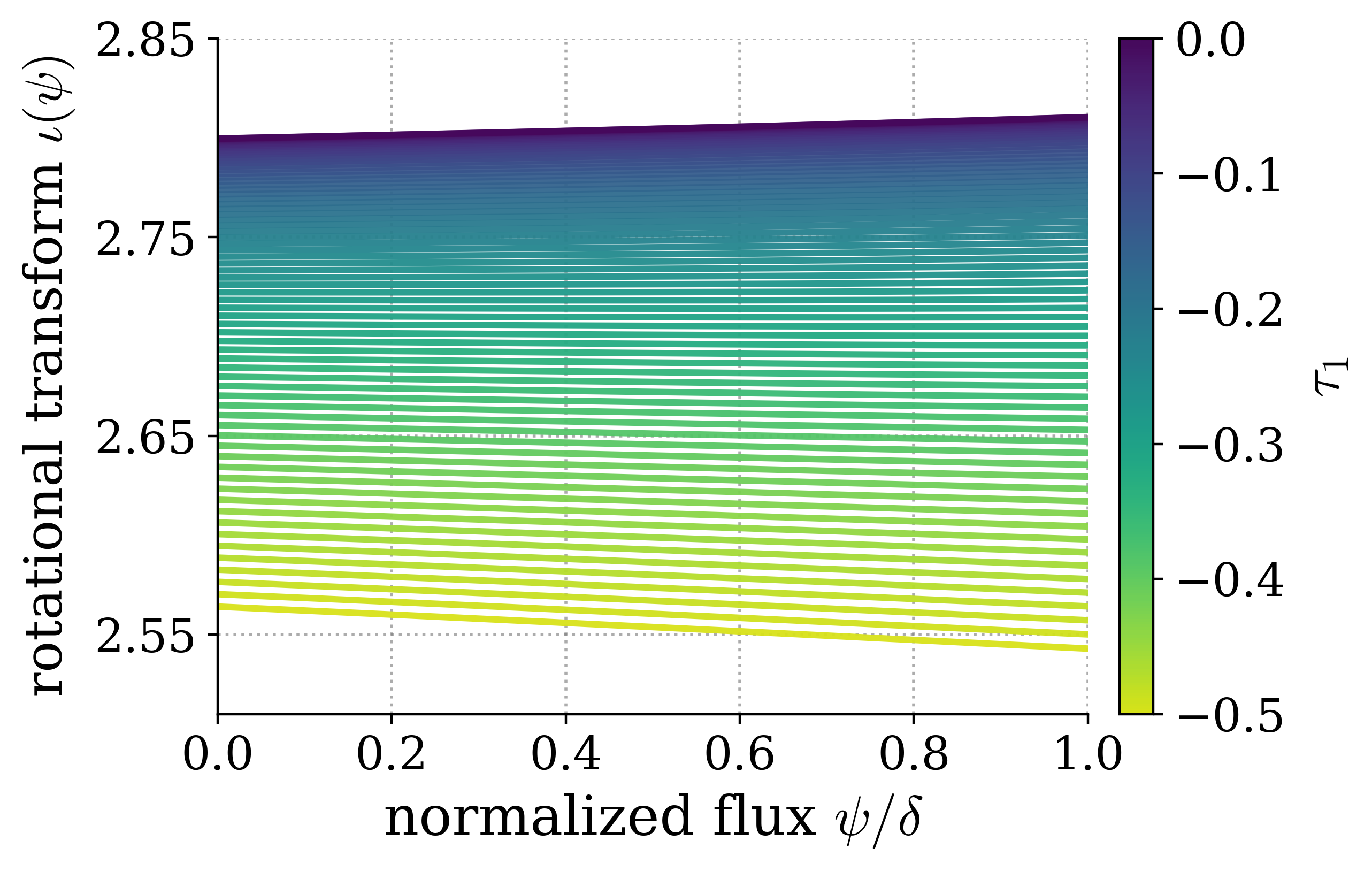}
    \label{fig:f2_iota_neg}
  \end{subfigure}
  \hfill
  \begin{subfigure}[b]{0.49\textwidth}
    \centering
    \caption{$\tau_{1} \ge 0$}
    \includegraphics[width=\textwidth]{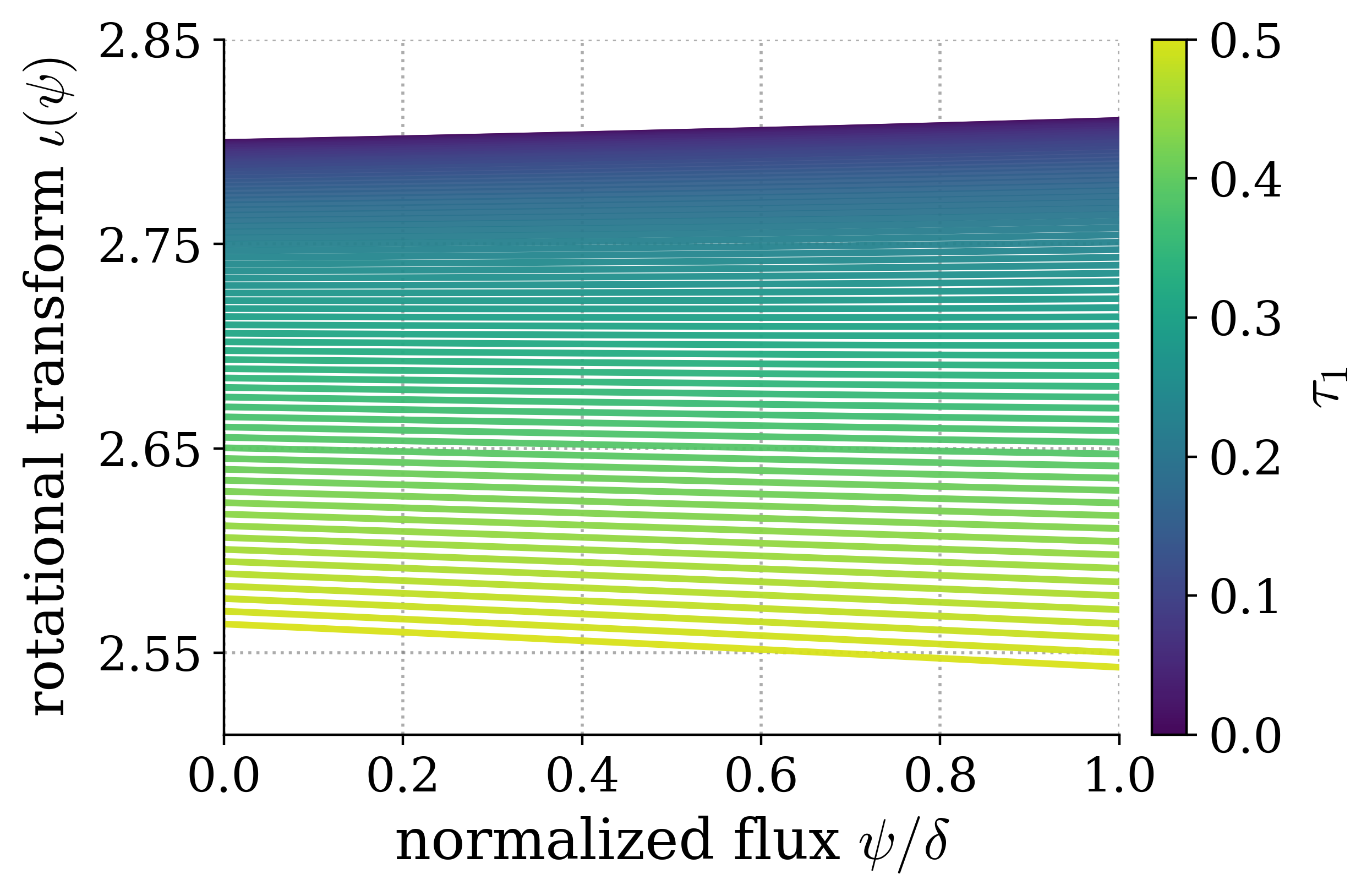}
    \label{fig:f2_iota_pos}
  \end{subfigure}
  \caption{Sheared rotational transform profile $\iota(\psi)$ as a function of normalized flux $\psi/\delta$ for  various values of $\tau_{1}$. Because $\iota(\psi)$ is an even function of $\tau_{1}$, the profiles for (a) $\tau_{1} \leq 0$ and (b) $\tau_{1} \geq 0$ are identical.}
  \label{fig:f2_iota}
\end{figure}

Parameterizing $(X, Y) = \sqrt{2\psi} (-\cos(\chi), \sin(\chi))$ in Eq.~\eqref{eq:f2_X_def} by the poloidal angle $\chi \in [0, 2\pi)$ as in Section~3.3 of~\citet{Landreman2026}, the rotational transform on a flux surface is given by
\begin{equation*} 
  \iota(\psi) \coloneq \lim_{N \to \infty} \frac{\chi(\zeta + 2\pi N) - \chi(\zeta)}{2\pi N},
\end{equation*}
where $\chi(\zeta)$ is obtained by integrating the field-line ordinary differential equation
\begin{equation} \label{eq:f2_dchi_dzeta}
  \frac{\mathrm{d}\chi}{\mathrm{d}\zeta} = \frac{\Bvec\cdot\grad\chi}{\Bvec\cdot\grad\zeta} = \big(h + \epsilon\cos(2\zeta)\big) \sqrt{\frac{1 - 2\psi\sin^2(\chi)}{\cosh^2(\tilde{\nu}) - 2\psi}}.
\end{equation}
Note that $\iota(\psi)$ depends only on $S$, $\epsilon$, and $\tau_{1}$ (as an even function of $\tau_{1}$), and is independent of $\lambda$.
Equation~\eqref{eq:f2_dchi_dzeta} does not admit a closed-form analytical solution, so we integrate it numerically for each $\psi$ and $\tau_{1}$ to obtain the profiles in Figure~\ref{fig:f2_iota} (with $\epsilon = 1.2$, $S = 1.4$, and $\delta = 0.14$).
Comparing Figures~\ref{fig:f2_iota_neg} and~\ref{fig:f2_iota_pos} shows that $\iota(\psi)$ is an even function of $\tau_{1}$ that decreases monotonically with $|\tau_{1}|$ while reversing the sign of the magnetic shear $\iota'(\psi)$ from positive near $\tau_{1} = 0$ to negative at larger $|\tau_{1}|$.

Moreover, differentiating $Y$ in Eq.~\eqref{eq:f2_X_def} along $\Bvec$ gives $\Bvec\cdot\grad\chi = \sqrt{1 - 2\psi\sin^2(\chi)}$, so along any field line $\mathrm{d}\ell/|\Bvec| = \mathrm{d}\chi/(\Bvec\cdot\grad\chi)$. On any rational flux surface $\iota(\psi) = m/n$, where closed field lines complete $m$ poloidal turns ($\Delta\chi = 2\pi m$) in $n$ toroidal turns, periodicity of $\sin^2(\chi)$ yields
\begin{equation*}
  \oint \frac{\mathrm{d}\ell}{|\Bvec|} = \int_{\chi_0}^{\chi_0 + 2\pi m} \frac{\mathrm{d}\chi}{\sqrt{1 - 2\psi\sin^2(\chi)}} = m \int_0^{2\pi} \frac{\mathrm{d}\chi}{\sqrt{1 - 2\psi\sin^2(\chi)}},
\end{equation*}
which is a complete elliptic integral of the first kind that depends only on $\psi$ and $m$ (and is independent of the initial field-line angle $\chi_0$, as well as $\tau_1$, $\epsilon$, $S$, and $\lambda$), confirming that the Hamada condition~\cite{Hamada1962} is satisfied on all rational surfaces.

\subsection{Enclosed volume and volume-averaged \texorpdfstring{$\beta$}{beta}}
\label{subsec:f2_vol_beta}
Following Section~3.4 of~\citet{Landreman2026}, we define 
\begin{align*} 
  \mathcal{I}[f] &\coloneq \int_0^{2\pi} \int_{-\xi_b}^{\xi_b} f(\xi, \zeta) \mathrm{d}\xi \mathrm{d}\zeta, \Hquad U \coloneq \arcsin\left(\sqrt{\frac{2\delta - \sin^2(\xi)}{\cosh^2(\tilde{\nu}) - \sin^2(\xi)}}\right), \\
  \xi &\coloneq \sigma - S, \Hquad \xi_b \coloneq \arcsin(\sqrt{2\delta}).
\end{align*}
The enclosed volume and volume-averaged $\beta$ are given by
\begin{equation*}
  V(\delta) = \frac{\mathcal{I}[U]}{\lambda}, \qquad
  \beta_V = \frac{2\langle p \rangle_V}{\langle |\Bvec|^2 \rangle_V} = \frac{2\lambda^2 p_b + 2(\delta - \mathcal{M})}{\lambda^2 \mathcal{A} + \mathcal{C}},
\end{equation*}
where $p_b = p(\delta)$ is the boundary pressure and the dimensionless integrals $\mathcal{M} \coloneq \langle \psi \rangle_V$, $\mathcal{C} \coloneq \langle X^2 \rangle_V$, and $\mathcal{A} \coloneq \langle B_x^2 + B_y^2 \rangle_V$ take the identical form to Section~3.4 of~\citet{Landreman2026} with $\nu$ replaced by $\tilde{\nu}$.

\begin{figure}[htbp]
  \centering
  \begin{subfigure}[t]{0.48\textwidth}
    \centering
    \caption{Enclosed volume $V(\delta)$}
    \includegraphics[width=\textwidth]{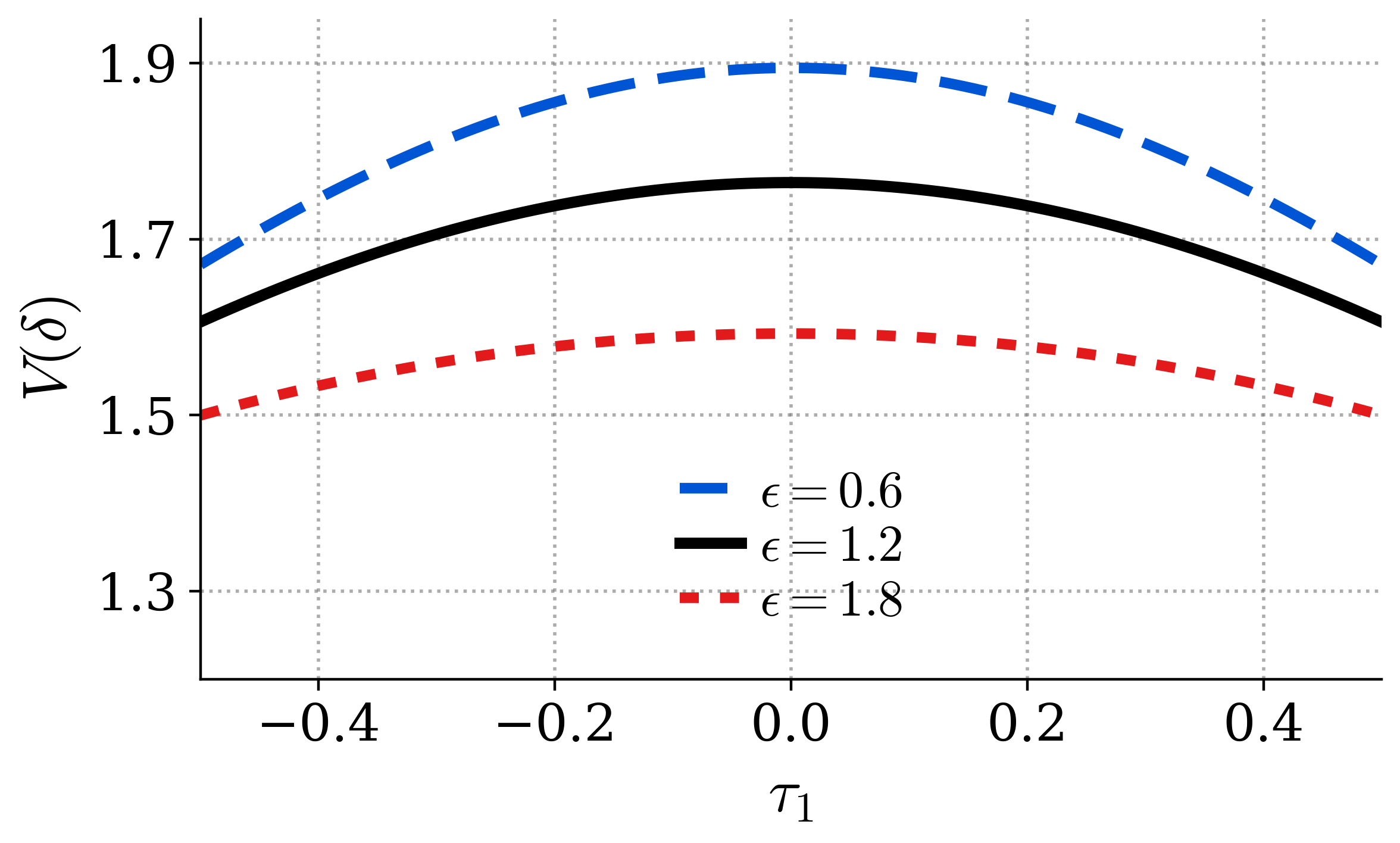}
  \end{subfigure}
  \hfill
  \begin{subfigure}[t]{0.48\textwidth}
    \centering
    \caption{Volume-averaged $\beta_V$}
    \includegraphics[width=\textwidth]{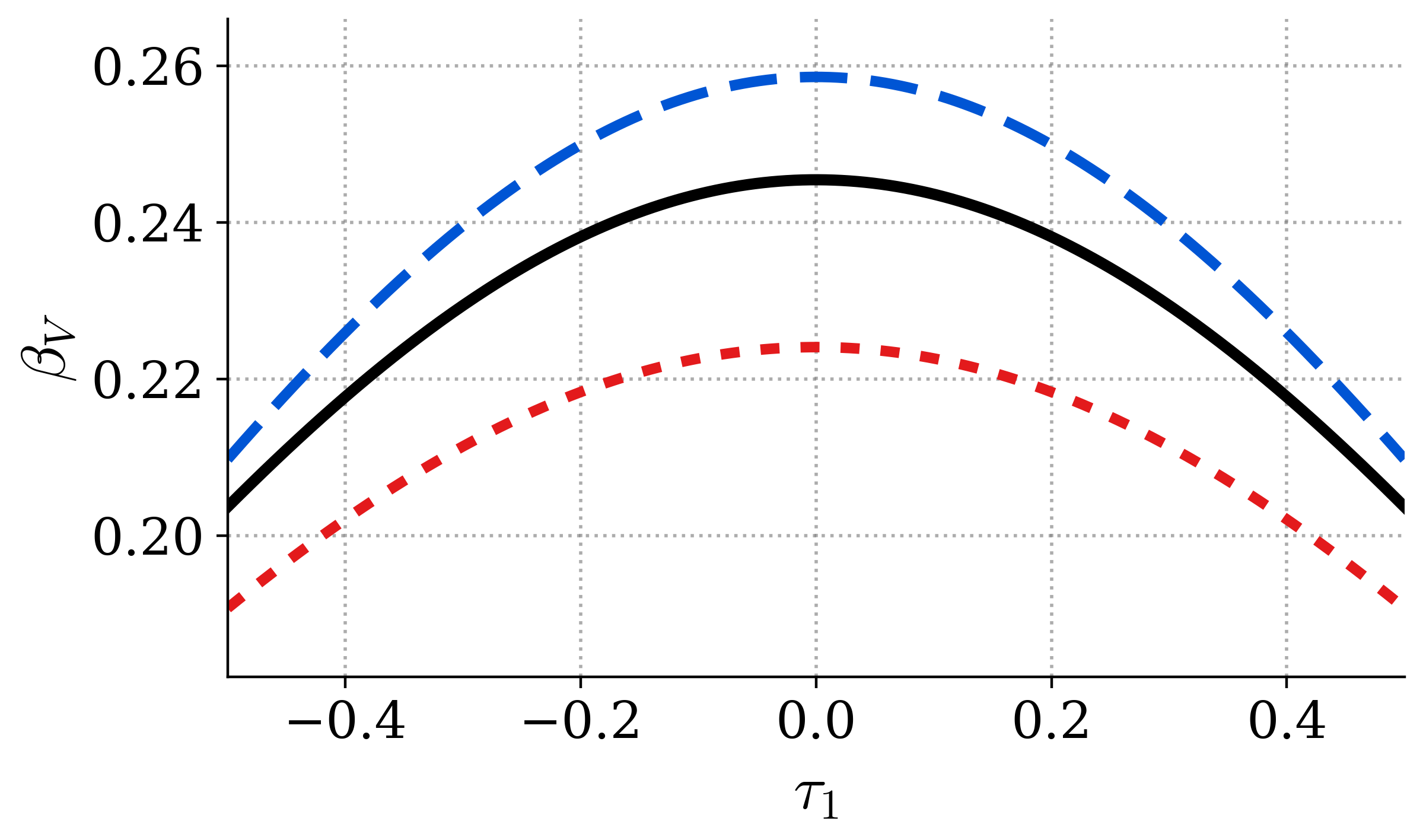}
  \end{subfigure}
  \caption{Enclosed volume $V(\delta)$ and volume-averaged plasma beta $\beta_V$ (with boundary pressure $p_b = 0$) vs.\ $\tau_{1}$ for the sheared-$\iota$ equilibrium family with $\delta = 0.14$, $S = 1.4$, and $\lambda = 1.6$. Both quantities are even in $\tau_{1}$ and decrease monotonically with both $|\tau_{1}|$ and $\epsilon$.}
  \label{fig:family2_vol_beta}
\end{figure}

As shown in Figure~\ref{fig:family2_vol_beta}, both $V(\delta)$ and $\beta_V$ are even functions of $\tau_{1}$, since reversing the sign of $\tau_{1}$ together with $\zeta \mapsto \pi - \zeta$ simply flips the sign of $\tilde{\nu}$ and leaves the integrands unchanged.
In contrast to the $\iota = 2$ family shown in Figure~\ref{fig:family1_vol_beta}, increasing $|\tau_{1}|$ reduces both the enclosed volume and $\beta_V$, because a larger $|\tau_{1}|$ increases $\cosh^2(\tilde{\nu})$, which compresses the flux-surface cross-sections and strengthens the magnetic field.

\subsection{Broken stellarator symmetry}
For $\tau_{1} \neq 0$, the equilibrium in Eqs.~\eqref{eq:f2_w_K_W_Xi_def}--\eqref{eq:f2_Q_pressure_psi} breaks the stellarator symmetry defined in Eq.~\eqref{eq:stellarator_symmetry}.
Under $\mathcal{S}\xvec = (x, -y, -z)$, the elliptic coordinates in Eq.~\eqref{eq:f2_xy} transform as $(\sigma, \zeta, z) \mapsto (\sigma, -\zeta, -z)$.
Evaluating $B_z(\mathcal{S}\xvec)$ from Eq.~\eqref{eq:f2_X_def} gives
\begin{equation*} 
\lambda B_z(\mathcal{S}\xvec) = -\sin(\sigma - S)\cosh(\tilde{\nu} - 2\tau_{1})\cos(\lambda z) - \cos(\sigma - S)\sinh(\tilde{\nu} - 2\tau_{1})\sin(\lambda z),
\end{equation*}
which differs from $B_z(\xvec)$ because of the arguments of the hyperbolic functions.

\begin{figure}[htbp]
  \centering
    \begin{subfigure}[t]{0.32\textwidth}
    \centering
    \caption{$\tau_{1} = 0$}
    \includegraphics[width=\textwidth]{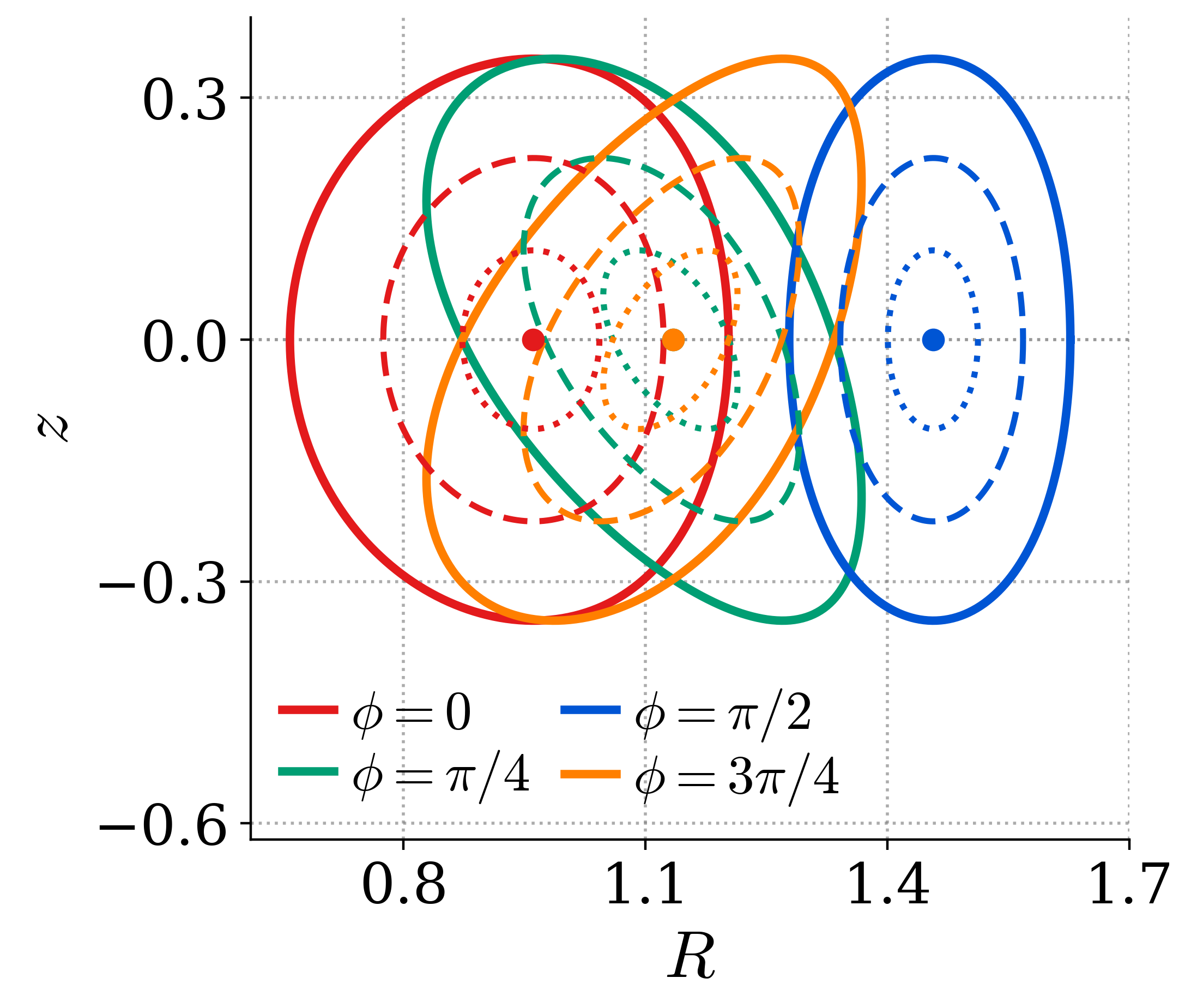}
    \label{fig:f2_rz_0}
  \end{subfigure}
  \begin{subfigure}[t]{0.32\textwidth}
    \centering
    \caption{$\tau_{1} = -0.40$}
    \includegraphics[width=\textwidth]{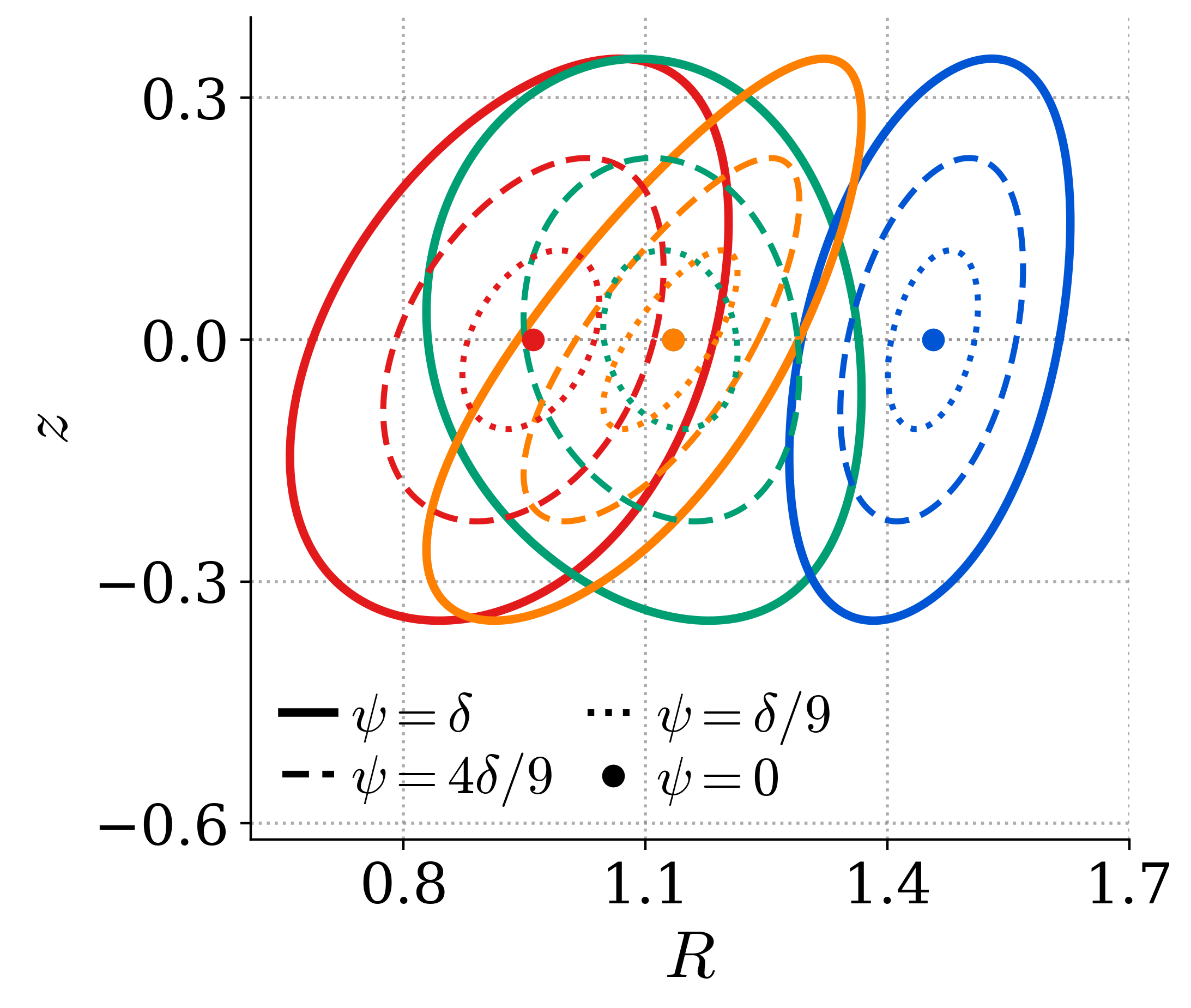}
    \label{fig:f2_rz_m040}
  \end{subfigure}
  \begin{subfigure}[t]{0.32\textwidth}
    \centering
    \caption{$\tau_{1} = 0.40$}
    \includegraphics[width=\textwidth]{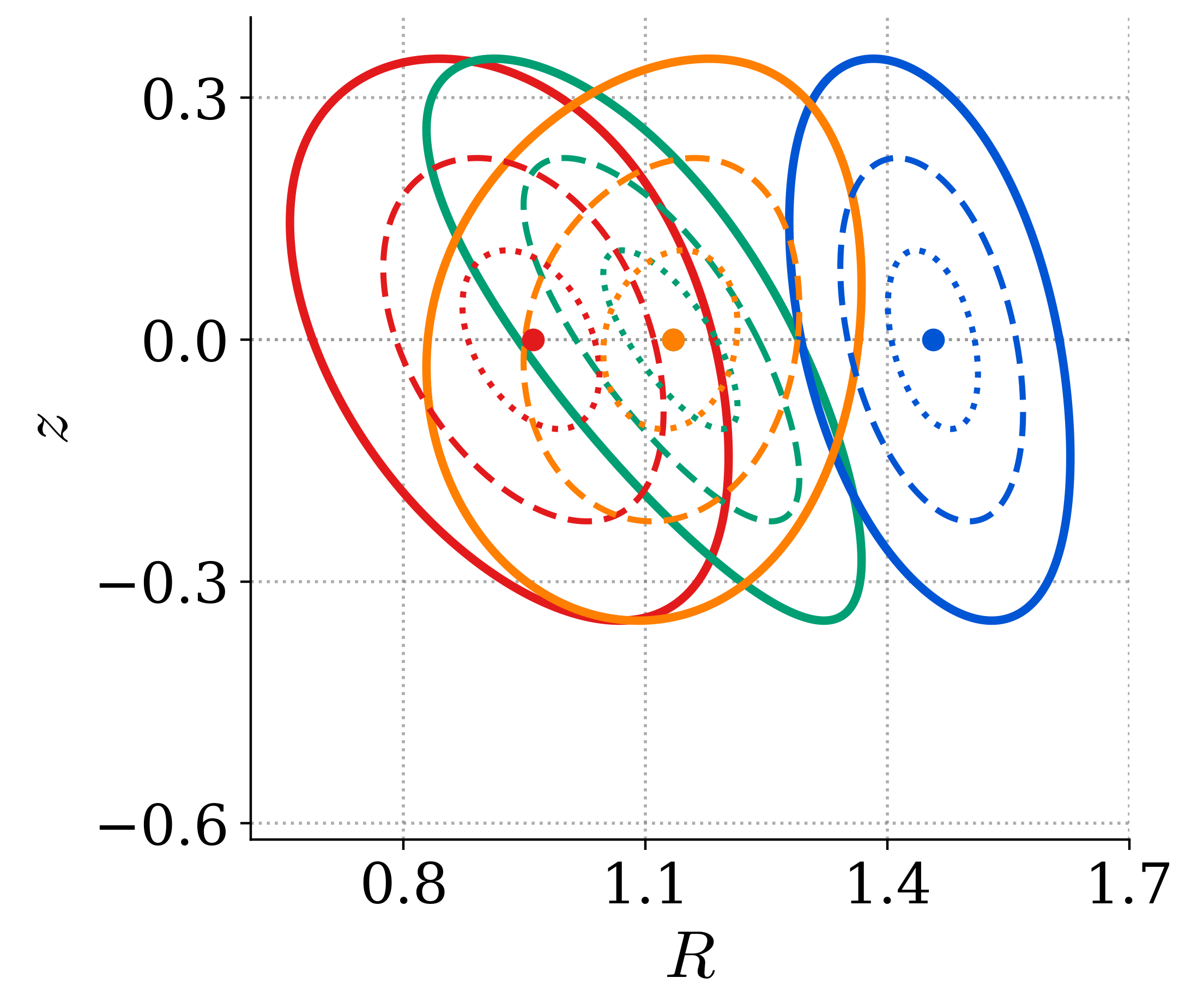}
    \label{fig:f2_rz_p040}
  \end{subfigure}
  \caption{Same as Figure~\ref{fig:family1_rz} for the sheared-$\iota$ equilibrium family across $\tau_{1}$. Subfigure~(a) with $\tau_{1} = 0$ is stellarator symmetric, with symmetric cross-sections at $\phi = 0, \pi/2$ and mirror-symmetric pairs at $\phi = \pi/4, 3\pi/4$, whereas nonzero $\tau_{1}$ in~(b) and~(c) breaks both symmetries.}
  \label{fig:family2_rz}
\end{figure}

Because $\tilde{\nu}(\zeta)$ is invariant under a half-turn $\zeta \mapsto \zeta + \pi$, the equilibrium retains $N_{\mathrm{fp}} = 2$ discrete rotational symmetry for all $\tau_{1}$, so the stellarator-symmetric case $\tau_{1} = 0$ obeys the cross-section symmetry in Eq.~\eqref{eq:f1_rz_symmetry}. This is shown in Figure~\ref{fig:f2_rz_0} with $\epsilon = 1.2$, $S = 1.4$, $\lambda = 1.6$, and $\delta = 0.14$.
As shown in Figures~\ref{fig:f2_rz_m040} and~\ref{fig:f2_rz_p040}, a nonzero $\tau_{1}$ leaves the magnetic axis fixed at $z = 0$ while tilting the $\phi = 0$ and $\phi = \pi/2$ cross-sections and breaking the vertical mirror symmetry between $\phi = \pi/4$ and $\phi = 3\pi/4$.

\section{Conclusions and future directions} \label{sec:summary}
We have presented two multi-parameter families of exact, smooth, non-axisymmetric three-dimensional ideal magnetohydrodynamic equilibria with nested toroidal flux surfaces and non-constant pressure that break stellarator symmetry and satisfy the Hamada condition. Both families generalize the stellarator-symmetric solutions recently discovered by~\citet{Landreman2026} through a single continuous symmetry-breaking parameter that recovers the corresponding stellarator-symmetric equilibrium when set to zero. 

In the constant-rotational-transform family of Section~\ref{sec:family1}, including an odd-in-$z$ term in the Solov'ev equilibrium~\cite{Solovev1968} prior to horizontal coordinate stretching deforms both the magnetic axis and the surrounding flux surfaces. In the elliptic family of Section~\ref{sec:family2}, introducing a complex phase shift leaves the planar elliptical magnetic axis unchanged while tilting and compressing the surrounding flux surfaces. Moreover, increasing the magnitude of this phase shift reduces the enclosed volume, volume-averaged plasma beta, and rotational transform. These analytical solutions provide additional closed-form reference benchmarks for verifying three-dimensional MHD equilibrium solvers across their full non-stellarator-symmetric Fourier solution spaces.

Beyond serving as extended verification benchmarks for three-dimensional equilibrium codes, these solutions demonstrate that smooth closed-form non-axisymmetric toroidal equilibria with finite rotational transform are not isolated artifacts of discrete stellarator symmetry. Future work could seek broader classes of exact equilibria.

\section{Use of AI tools} \label{sec:AI_agents}
This work was carried out using Gemini 4 running as a small team of agents in Google Antigravity, an agentic development environment. Given~\citet{Landreman2026} and a list of the open questions it raised, the agents proposed the generalized solutions that remove the stellarator-symmetry assumption. Promising candidates appeared within hours. Gemini 4 Argon was also used to draft parts of the manuscript and write symbolic and numerical verification code. The authors chose the problem, steered the agents, and ran several rounds of adversarial review. Every derivation was independently re-derived and checked symbolically by the authors. The authors refined the interpretation and rewrote the manuscript, and take full responsibility for its content.


\bibliographystyle{unsrtnat}
\bibliography{reference}

\end{document}